\documentclass[aps,prd,reprint,nofootinbib,longbibliography]{revtex4-2}
\usepackage[utf8]{inputenc}
\usepackage{graphicx,fancybox,float,comment}
\usepackage{units}
\usepackage{multirow,array}
\usepackage{xcolor}
\usepackage{amsmath,amssymb}
\usepackage{yfonts}
\usepackage{tensor}
\usepackage{paralist}
\usepackage{braket}
\usepackage{tensor,mathrsfs}
\usepackage[normalem]{ulem}

\newcommand{\para}{||}
\DeclareMathOperator\arctanh{arctanh}
\newcommand{\Left}{\textit{Left: }}
\newcommand{\Right}{\textit{Right: }}
\newcommand{\A}{\mathscr{A}}

\graphicspath{figs_Update}
\begin{document}
\title{
Hydrodynamic attractors for the speed of sound in holographic Bjorken flow
}

\newcommand{\uaAff}{\affiliation{Department of Physics and Astronomy, University of Alabama, 514 University Boulevard, Tuscaloosa, AL 35487, USA}}
\newcommand{\uuAff}{\affiliation{Institute for Theoretical Physics, Utrecht University, Princetonplein 5, 3584 CC Utrecht, The Netherlands}}
\newcommand{\Say}[1]{``#1"}
\author{Casey Cartwright}
\email{c.c.cartwright@uu.nl}
\uaAff
\uuAff
\author{Matthias Kaminski}
\email{mski@ua.edu}
\author{Marco Knipfer}
\email{mknipfer@crimson.ua.edu}
\uaAff
\date{\today}

\newcommand{\dd}{\mathrm{d}}
\newcommand{\vev}[1]{\braket{#1}}

\begin{abstract}
The time evolution of the averaged energy momentum tensor as well as its variation with energy density are calculated in a far-from-equilibrium state of $\mathcal{N}=4$ SYM theory undergoing a Bjorken expansion.
The calculation is carried out holographically where we consider a collection of trajectories of the energy density in the space of solutions by small changes to the initial conditions of the bulk spacetime. 
We argue that the proper interpretation of the variation of the diagonal energy momentum tensor components with respect to the energy density is that of a far-from-equilibrium speed of sound. We demonstrate remarkable agreement with a corresponding hydrodynamic prediction. 
We find by Borel resummation that the holographic system has one attractor for this speed of sound longitudinal, and another transverse to the direction of Bjorken expansion.
Attractor times for various initial flow conditions show that reaching an attractor does not imply or require local thermal equilibrium. In the cases studied, reaching an attractor implies hydrodynamization (quantities evolve approximately according to hydrodynamics), justifying the name \emph{hydrodynamic attractor }. 
\end{abstract}

\maketitle

\section{Introduction} 
\label{sec:intro}

One of the basic hydrodynamic assumptions---like in all effective field theories---is that the contributions from
derivatives of operators are less important than the operators themselves, this leads to the gradient expansion of conserved quantities like the energy-momentum tensor. 
Surprisingly, hydrodynamics works well for the description of heavy ion collisions already early after the collision, where gradients are still expected to be large. 
Another way of saying this is that the hydrodynamic expansion is like an expansion in the Knudsen
number and already works well for times when the Knudsen number is still large~\cite{noronha-hostlerUnreasonableEffectivenessHydrodynamics2016}. This \Say{unreasonable effectiveness of hydrodynamics}~\cite{noronha-hostlerUnreasonableEffectivenessHydrodynamics2016} might be explained by the presence of \emph{hydrodynamic attractors}, which implies that after a rather short time the initial deviations from a hydrodynamic evolution die away exponentially fast in a holographic strongly coupled system~\cite{Kurkela:2019set}\footnote{ AdS/CFT models for strongly coupled $\mathcal{N}=4$ SYM theory display no distinct early-time attractor~\cite{Romatschke:2017vte,Kurkela:2019set}, while in Israel-Stewart and kinetic theory the universal attractor extends to arbitrarily early times. Therein, the approach to the attractor at early times is governed by a power-law (presumably driven by the expansion of the plasma) and it is exponential at late times (presumably driven by collisions).} and the system follows the hydrodynamic evolution independent from the exact initial conditions. Naturally, attractor behavior has also been seen in non-holographic contexts as well, such as those found within QCD and kinetic theory~\cite{Heller:2016rtz,Denicol:2019lio,Almaalol:2020rnu,Du:2022bel}. Anisotropic attractors were considered~\cite{Strickland:2017kux}, and attractors (including early-time attractors at weak coupling) have been further studied in the context of Bjorken flow with higher-order viscous fluid dynamics~\cite{Jaiswal:2019cju} (also for Gubser flow~\cite{Chattopadhyay:2018apf}), non-conformal systems~\cite{Chattopadhyay:2021ive}, and in non-conformal kinetic theory~\cite{Jaiswal:2021uvv}, see also~\cite{Chattopadhyay:2019jqj}.\footnote{These non-conformal systems show a progressing destruction of the attractor behavior (depending on coupling strength and on the degree to which the conformal symmetry is broken). It is intriguing that the longitudinal pressure, $P_L$, can still have an early-time attractor~\cite{Chattopadhyay:2021ive}. However, see the more recent results from~\cite{Kamata:2022jrc}.
}

Given that hydrodynamic models display attracting behavior of the one point functions of the energy momentum tensor, one may ask themselves if higher point correlations also display this behavior. It is our goal, here, to initiate such a study by continuing the story of attracting behavior of the boost invariant evolution in $\mathcal{N}=4$ SYM plasma. 
Using the characteristic formulation of general relativity, we obtain numerical solutions to the Einstein equations. These correspond to the out-of-equilibrium evolution of data, which may be thought of as initialized shortly after the ``collision'' of two heavy ions.
To simulate correlations with our evolutions we study the variation of the energy momentum tensor with respect to the energy density. We conduct this study by considering a class of initial conditions sightly varied around a central initial condition. 
We argue that the variation of the energy momentum tensor, as a result of these perturbations of initial conditions, reflects a potential out-of-equilibrium speed of sound.
Utilizing the confirmed hydrodynamic attractors of $\mathcal{N}=4$ SYM theory we provide a leading order resummation of the hydrodynamic expectation of the speed of sound, and find excellent agreement between the exact numerical evolution and the hydrodynamic attractor expectation. 

We note that due to the anisotropy of Bjorken flow, there are two distinct derivatives with respect to the energy density, because the energy-momentum tensor has diagonal components longitudinal ($\langle T_{||\,||}\rangle$) and transverse ($\langle T_{\perp\perp}\rangle$) to the direction of the expansion. 
We compute the hydrodynamic value for these derivatives and compare these expectations to the same derivatives computed with our numerical far-from-equilibrium evolutions, finding remarkable agreement from early times onward. 
In the discussion section we propose these two derivatives as the two distinct speeds at which sound waves propagate through the Bjorken expanding plasma longitudinal versus transverse to the expansion.\footnote{Sound modes on top of Bjorken flow were already studied in kinetic theory~\cite{Kurkela:2018vqr,Kamata:2020mka}.}

In addition, using our numerical evolutions, we take the opportunity to clarify previous results on entropy production in holographic models~\cite{Rougemont:2021qyk} and confirm previous numerical results about the hydrodynamic attractor of $\mathcal{N}=4$ SYM theory~\cite{Heller:2015dha,spalinskiHydrodynamicAttractorYangMills2018}. We further note that one way to interpret the results of~\cite{Kurkela:2019set} is that in holographic systems hydrodynamization occurs at the same time scale as the system reaches the hydrodynamic attractor, a point we confirm in the discussion of our results, relating it to the time scale at which local thermal equilibrium is reached, see Fig.~\ref{fig:attraction_times}.

We begin our work with a brief review of hydrodynamics and the symmetries of a boost invariant plasma in section~\ref{sec:BjorkenFlow}. We then introduce the holographic model with which we will work in section~\ref{sec:setup}. Following this we discuss the calculation of the speed of sound in thermodynamic systems and its extension to hydrodynamic evolution in section~\ref{sec:SpeedOfSound}. Here we compare our results with the out-of-equilibrium gravitational calculation and discuss them in the context of other out-of-equilibrium thermodynamic quantities such as the entropy. We conclude this section with a verification of the hydrodynamic attractor $\mathcal{N}=4$ SYM theory as well as a derivation of expressions for the resummed speeds of sound. Finally we conclude our work in section~\ref{sec:conclusions} with some discussion and questions to be investigate in future work.

\section{Hydrodynamics}
\label{sec:BjorkenFlow}
The modern view on hydrodynamics is that it is the \textit{long-wavelength} effective theory of some microscopic theory. Hydrodynamics is a field theory of conserved quantities which are conserved due to symmetries. 
The hydrodynamic fields are
\begin{itemize}
    \item the \textit{fluid velocity} $u^\mu(x)$,
    \item the \textit{temperature} $T(x)$,
    \item possibly other fields if charges etc.\ are added.
\end{itemize}
Unlike in quantum field theory, in hydrodynamics one does not start from a generating functional
$\Gamma\left[ u^\mu, T, \ldots, \partial_\alpha u^\mu, \partial_\alpha T, \ldots \right]$,
but from the 1-point functions of conserved currents.\footnote{See~\cite{jensenHydrodynamicsEntropyCurrent2012, banerjeeConstraintsFluidDynamics2012,Jensen:2011xb} for the first constructions of generating functionals for hydrodynamics, and the first frameworks for including dissipative terms in a hydrodynamic generating functional~\cite{Haehl:2015pja,Haehl:2015foa,Crossley:2015evo}, for an accessible review see~\cite{Glorioso:2018wxw}.
}  
For example, an ideal hydrodynamic description of an uncharged fluid has conserved 
currents consisting only of the \textit{energy-momentum tensor}
\begin{equation}
    \langle T^{\mu\nu}_{(0)} \rangle = (\epsilon + P)u^\mu u^\nu + P g^{\mu\nu}\,,
    \label{eq:Tmunu0}
\end{equation}
which is usually just written as $T^{\mu\nu}_{(0)}$ (without the $\langle \cdot \rangle$).
The energy-momentum tensor is conserved
\begin{equation}
    \nabla_\mu T^{\mu\nu}_{(0)} = 0\,,
\end{equation}
where $\nabla_\mu$ is the covariant derivative (simply $\partial_\mu$ in Minkowski space-time) and
this equation is also called the \textit{relativistic Euler equation}.
If conserved charges are present, say an electric charge $\rho(x)$, then a conserved current at leading order in
the derivative expansion would have the form $j_\mu = u_\mu \rho(x)$.
The form the currents take as a function of the fields is named \textit{constitutive equations}.
Generally, also an \textit{equation of state} (EOS) is needed and often it is given\footnote{Examples are \textit{dust}, $P(\epsilon) = 0$, and \textit{(conformal) relativistic matter} in 3+1 dimensions, $P(\epsilon) = \epsilon/3$.} in the form
$P(\epsilon)$.

\begin{figure}
    \centering
    \includegraphics[width=0.4\textwidth]{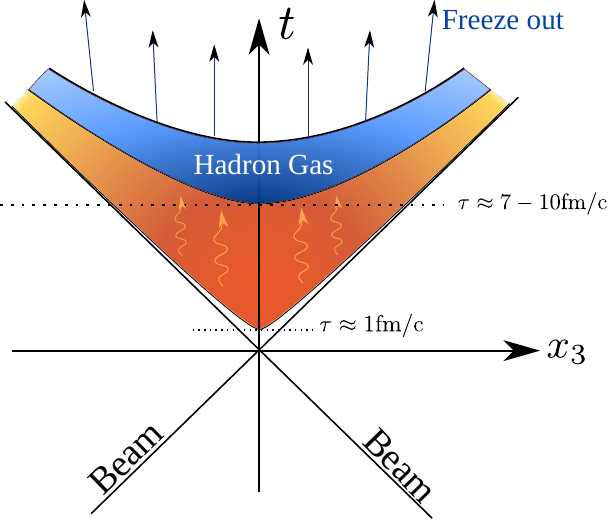}
    \caption{Schematic sketch of \textit{boost invariance} along the beam axis, which is assumed by Bjorken invariance.
    The hyperbolas are lines of constant proper time $\tau$ and the time evolution only depends on $\tau$, the system is invariant
    along the hyperbolas.
    } 
    \label{fig:BjorkenFlow}
\end{figure}

Symmetries play an important role in reducing the complexity of hydrodynamic equations. One highly symmetric flow, in particular, has been of enormous use in understanding the hydrodynamic evolution of heavy ion collisions: boost invariant expansion. While studied previously, it is in 1982, that James D.\ Bjorken published a seminal study of the time evolution of the central region of heavy ion collisions~\cite{Bjorken:1982qr}. In this work, central collisions\footnote{Ignoring the spectator nucleons is a simplification that might be much too restricting.
One can imagine that an off-central collision would lead to rotation because the spectator nucleons
would drag the resulting lump into a rotation.
It has also been measured that heavy ion collisions lead to the most vortical fluid~\cite{STAR:2017ckg}.
Progress is being made on the spinning case based on holography~\cite{bantilanHolographicHeavyIonCollisions2018, garbisoHydrodynamicsSimplySpinning2020,Cartwright:2021qpp}.
}
of large nuclei are considered, where for transverse distances much smaller than the nuclear radii, the fluid expansion of QGP near the collision axis is longitudinal and homogeneous\footnote{As stated in~\cite{Bjorken:1982qr}, for distances on the order of the nuclear radii there is a rarefaction front moving inward towards the central region at the speed of sound of the medium. While for distances larger than this, the fluid expands radially outwards.}. 
In the longitudinal\footnote{Within this region the fluid may be considered to have a translational and rotational invariance in the plane transverse to the beam direction. A generalization of this longitudinal flow that allows for transverse expansion is \textit{Gubser flow}~\cite{Gubser:2010ze}.} direction the fluid, a distance $z$ from the stationary center, moves along the beam direction with longitudinal velocity $v/t$, where $t$ is the time elapsed since the collision occurred. The most important assumption of this work is \Say{the existence of a central plateau structure for the particle production as a function of rapidity.} This assumption implies that boosts, with $\gamma$ much smaller than that of the colliding nuclei, do not affect the description of the fluid and hence that the initial conditions for the fluid are the same as those that existed in any other Lorentz frame, i.e.\ the physics of the longitudinal expansion depends only on the spacetime interval $\dd s^2=\sqrt{\dd t^2-\dd x_3^2}$. This leads naturally to the description of the longitudinal expansion in terms of proper time $\tau=\sqrt{t^2 - x_3^2}$ and spacetime rapidity $\xi = \arctanh{x_3/t}$. This situation is depicted in Figure~\ref{fig:BjorkenFlow} (where the transverse directions have been suppressed). Incoming from the left and right sides, the two beams collide at the origin. The hyperbola shaped lines are lines at constant $\tau$. Bjorken's assumption of a central plateau implies that the evolution only depends on $\tau$, thus along each hyperbola the system looks the same. 
The solutions obtained for the hydrodynamic equations under these assumptions\footnote{Note, that by \Say{solution} here we also refer to quantities which may only be known in terms of an asymptotic series expansion, e.g., in $\tau$~\cite{Heller:2015dha}.} are particularly simple, especially so when ignoring viscous effects. Remarkably, despite the simplicity of the solutions and the rather restricting assumptions, practice shows that the flow discussed in~\cite{Bjorken:1982qr} provides reliable results, 
consistent with the experimental measurements~\cite{Romatschke:2017ejr}. 

Let us now consider, briefly, some details associated with the second order gradient expansion of the one-dimensional longitudinal expansion as described in~\cite{Bjorken:1982qr}. To second order in the gradient expansion we include an additional term\footnote{We can in fact add another term to the energy-momentum tensor at first order $\zeta\Delta^{\mu\nu}  \Delta_{\lambda \sigma} \nabla^{\sigma} u^{\lambda}$ where $\zeta$ is the bulk viscosity. However for a conformal fluid this contribution must vanish to preserve the trace condition on the energy-momentum tensor. } in the energy-momentum tensor
\begin{equation}
    T^{\mu\nu}=(\epsilon+P)u^{\mu}u^{\nu}-Pg^{\mu\nu}+\pi^{\mu\nu} \, ,
\end{equation}
 where $\pi$ is symmetric, traceless ($\pi^\mu_\mu=0$) and $\partial_{\mu}\pi^{\mu\nu}=0$. This is the shear stress tensor and to second order it is given by
\begin{align}
   \pi^{\mu\nu} ={}&-\eta\sigma^{\mu\nu}+\eta\tau_\pi \left(^<u^\alpha \nabla_\alpha\sigma^{\mu\nu>}+\frac{\nabla^\alpha u_\alpha}{d-1}\sigma^{\mu\nu}\right) \nonumber \\ &+\kappa \left(R^{<\mu\nu>}-2u_\lambda u_\rho R^{\lambda <\mu\nu>\rho}\right) \nonumber\\ 
    & +\lambda_1\tensor{\sigma}{^{<\mu}_\lambda}\sigma^{\nu>\lambda}+\lambda_2\tensor{\sigma}{^{<\mu}_\lambda}\Omega^{\nu>\lambda}+\lambda_3\tensor{\Omega}{^{<\mu}_\lambda}\Omega^{\nu>\lambda} \, , \label{eq:shear_2nd_order}
\end{align}
where $R$ is the curvature tensor (4 index) or the Ricci tensor (2 index) and $\Omega$ is the fluid vorticity. The quantity $\sigma$ is defined as
\begin{equation}
 \frac{1}{2}\sigma^{\mu\nu}=\partial^{<\mu} u^{\nu>}
\end{equation}
where the projector is defined as $\Delta^{\alpha\beta}=u^\mu u^\nu -g^{\mu\nu}$ and we have introduced the notation for the projection operation defined as~\cite{Romatschke:2017ejr}
\begin{equation}
B^{<\mu\nu>}=\frac{1}{2}\Delta^{\mu\alpha}\Delta^{\nu\beta}(B_{(\alpha\beta)})-\frac{1}{d-1}\Delta^{\mu\nu}\Delta^{\alpha\beta}B_{\alpha\beta} \, ,
\end{equation}
for a general rank 2 tensor.

For the case at hand, both the Riemann and Ricci tensor and the fluid vorticity vanish. The resulting expression for the stress tensor is given by
\begin{equation}
   \pi^{\mu\nu} =\bar{\pi}\text{diag}\left(0,1,1,-2\right)\, , \quad
   \bar{\pi}=\frac{2 \eta }{3 \tau }-\frac{4 \lambda_1}{9 \tau ^2}+\frac{4 \eta  \tau_\pi}{9 \tau ^2}\, .
\end{equation}
We then see that for a conformal fluid ($P = \epsilon/3$) the viscous hydrodynamic equations in a boost invariant flow to second order in the expansion\footnote{This expression is nonlinear in amplitudes, an expansion in increasing number of gradients, including dissipative effects.} in proper time $\tau$ become~\cite{Baier:2007ix, Romatschke:2017ejr}
\begin{equation}
    \partial_{\tau} \epsilon+\frac{4\epsilon}{3\tau}=\frac{4 \eta}{3 \tau^{2}}+\frac{8 \eta \tau_{\pi}}{9 \tau^{3}}-\frac{8 \lambda_{1}}{9 \tau^{3}}\,,
    \label{eq:BjorkenEpsilon}
\end{equation}
where $\eta$, $\tau_\pi$ are first order transport coefficients, namely the shear viscosity and relaxation time  respectively, and $\lambda_1$ is a second order transport coefficient. Scaling under conformal transformations allows a redefinition of the transport coefficients, equation~(\ref{eq:BjorkenEpsilon}) can be written as~\cite{Baier:2007ix, Romatschke:2017ejr}
\begin{equation}
    \tau \partial_{\tau} \ln \epsilon=-\frac{4}{3}+\frac{16 C_{\eta}}{9 \tau T}+\frac{32 C_{\eta} C_{\pi}\left(1-C_{\lambda}\right)}{27 \tau^{2} T^{2}}\,,
    \label{eq:BjorkenEpsilon2}
\end{equation}
where $T(\tau) = (\epsilon(\tau) / \sigma_\text{SB})^{1/4}$ can be interpreted as temperature. For $\mathcal{N}=4$ SYM the dimensionless transport coefficients take the values~\cite{Romatschke:2017vte}
\begin{equation}\label{eq:transport}
    C_\eta=\frac{1}{4\pi}\, ,\quad C_\pi=\frac{2-\log(2)}{2\pi}\, , \quad C_\lambda =\frac{1}{2-\log(2)} \, .
\end{equation}
In terms of these coefficients we can solve for the temperature from Eq. (\ref{eq:BjorkenEpsilon2}), which to third order in $\tau$ is given as
\begin{align}
   T&= \frac{\tilde{\Lambda}}{(\tilde{\Lambda}  \tau)^{1/3}}  \left(1-\frac{1}{6 \pi  (\tilde{\Lambda}  \tau)^{2/3}}+\frac{\log (2)-1}{36 \pi ^2 (\tilde{\Lambda}  \tau)^{4/3}} \right. \nonumber  \\
    &\left.+\frac{2 \pi ^2-21-24 \log ^2(2)+51 \log (2)}{1944 \pi ^3 (\tilde{\Lambda}  \tau)^2}\right) \, .
\end{align}
We will refer to this solution for the temperature as $T_{3rd}$. We will use these hydrodynamic approximations for comparison to our numerical data below.

\section{Holographic setup \& numerics}
\label{sec:setup}
As a gravitational dual theory, we consider the Einstein-Hilbert action 
\begin{equation}
    S=\frac{1}{16\pi G}\int \dd^5x\, \sqrt{-g}(R-2\Lambda) 
\end{equation}
for which $G$ is the five dimensional Newton constant, and the cosmological constant is given in terms of the AdS radius $L$, by $\Lambda=-6/L^2$. The numerical technique we work with was pioneered by Chesler and Yaffe~\cite{Chesler:2008hg} (an excellent review is given in~\cite{Chesler:2013lia}). 
We now discuss the general method for solving the Einstein Field Equations (EFEs) via the characteristic method. We begin by fixing a general metric ansatz in generalized Eddington-Finkelstein coordinates
\begin{equation}\label{eq:Gen_Line_Element}
    \dd s^2=2\dd r\dd v - \frac{r^2}{L^2}g_{\mu\nu}\dd x^{\mu} \dd x^{\nu} \, .
\end{equation}
A further reduction of this ansatz consistent with the symmetries of the system is given in the next section. Given an ansatz we begin by writing the EFEs in the \textit{characteristic formulation} using directional derivatives referred to as characteristic derivatives 
\begin{equation}
    \dot{\Phi}(v,r)=\partial_v\Phi+\frac{1}{2}g_{00}\partial_r\Phi \, .
\end{equation}
These directional derivatives point along out-going null geodesics in generalized infalling Eddington-Finkelstein coordinates. The foliation of the spacetime into null hypersurfaces in this way leads to the EFEs developing a nested structure. Starting from some initial data on an initial time slice $v_0$, the EFEs can be solved to obtain the full metric at this time. From the definition of the dotted derivative and from the boundary expansion one can obtain the time evolution equations required to propagate the initial data to the next time slice. The procedure is the then repeated until a final time slice is reached. 

The time evolution itself can be written schematically as
\begin{equation}
    \frac{\dd \Phi}{\dd t} = \mathcal{F}[\Phi]\, ,
\end{equation}
where $\mathcal{F}[\Phi]$ can be complicated to calculate. To obtain this one has to go through the nested system of differential equations.
Then, given an initial $\Phi(v_0)$, the data can be propagated to the next time slice using one's favorite time stepping algorithm.

This procedure is not new, it has been used in a large number of publications (see for example~\cite{Chesler:2008hg,Chesler:2009cy,Chesler:2010bi,Chesler:2013lia,Cartwright:2019opv,Cartwright:2020qov,Cartwright:2021maz}). For this reason we relegate a large portion of the details associated with our numerical solutions to appendix~\ref{appendix:Numerics}. In the following we will only give basic details required for the remaining exposition of this work. 

\subsection{Metric ansatz}
Consistent with the symmetries discussed in section~\ref{sec:BjorkenFlow} the metric ansatz given in Eq.~(\ref{eq:Gen_Line_Element}) can be reduced to 
\begin{align}
    \dd s^2&=2 \dd r\dd v -A(v,r)\dd v^2+e^{B(v,r)}S(v,r)^2(\dd x_1^2+\dd x_2^2)\nonumber \\
    &+S(v,r)^2 e^{-2B(v,r)}\dd \xi^2\,,\label{eq:line_element}
\end{align}  
where $v$ is the Eddington-Finkelstein time, $r$ is the bulk AdS direction, $x_1$ and $x_2$ are the coordinates in the plane transverse to the beamline and $\xi=\frac{1}{2} \ln [(t+x_3)/(t-x_3)]$ is the rapidity in longitudinal direction. As discussed in section~\ref{sec:BjorkenFlow} the conservation equations of the fluid at the conformal boundary of AdS spacetime will depend only on $\tau$. With that in mind, it is useful to set the boundary metric in terms of the coordinates $(\tau,x_1,x_2,\xi)$ to be given by 
\begin{equation}
    \lim_{r\to \infty} \frac{1}{r^2}\dd s^2=-\dd \tau^2+\dd x_1^2+\dd x_2^2+\tau^2\dd \xi^2 \, .
\end{equation}
Comparing to equation~(\ref{eq:line_element}), this places boundary conditions on the metric functions $A, B, S$, namely:
\begin{align}
  \lim_{r\to \infty}  A&\to r^2\,, \\
  \lim_{r\to \infty}  B&\to \log \left(\frac{1}{\tau^{2/3}}\right)\,,\\ 
  \lim_{r\to \infty}    S&\to r \tau^{1/3}\,.
\end{align}
Also, $\lim_{r\to\infty} v = \tau$, so the Eddington-Finkelstein time is the proper time on the boundary.

\subsection{Initial conditions}
The initial data required to begin the evolution consists an initial time $v_0$, an initial value of the asymptotic coefficient $a_4$ (dual to the energy density), an initial value for the radial shift diffeomorphism $\lambda$ as well as a profile for $B(z,v_0)$ on the initial time slice (where we have already changed variables, $r=1/z$, placing the boundary at the finite location $z=0$). For the initial profile we follow previous authors~\cite{Kurkela:2019set,Rougemont:2021qyk,Cartwright:2021maz} and choose to parameterize our solutions as deviations away from a vacuum $AdS$ solution to the Einstein equations.
\begin{equation}\label{eq:init_dev}
    B = B_\text{d} + B_\text{AdS}
\end{equation}
Where one can check directly that $B=B_\text{AdS}=-2/3\log(v+z)$ is a solution to the Einstein equations provided $A=A_\text{AdS}=z^{-2}$ and $S=S_\text{AdS}=z^{-2/3}(1 +  v/z)^{1/3}$. 

To implement this choice of parameterization of the initial profile one has to be careful to merge this choice with the choice made of the parameterization of the function $B$ used to construct the numerical routine. There one works with ``subtracted'' functions, defined to remove singular terms from the function. This is done since our choice of spectral decomposition is only well suited to the approximation of regular functions. Hence we work with the following schematic form (the exact scheme we work with is given in Eq.~(\ref{eq:SubtractionScheme})) of the metric components in our numerical scheme,
\begin{align}\label{eq:redef}
    B &= z^4 B_\text{s} + \Delta_B \, ,\\
    S &= z^4 S_\text{s} + \Delta_S \, , \\
    A &=  A_\text{s} + \Delta_A \, .
\end{align}
To optimize the routine one analytically inserts the redefinitions given in Eq.~(\ref{eq:redef}) into the Einstein equations and simplifies the resulting equations. Doing so leads to the equations of motion being written for the regular functions $B_\text{s},S_\text{s}$ and $A_\text{s}$ rather than the singular functions $B,S$ and $A$. 

Given that the equations of motion are now written in terms of the regular functions, rather than the singular functions, the initial data actually required to begin the evolution is for the regular function $B_\text{s}(u,v_0)$ on the initial time slice. This requires us to translate the data prescribed in Eq.~(\ref{eq:init_dev}) as,
\begin{equation}\label{eq:init_data}
    B_\text{s}=\frac{1}{z^4}\left(
     B_\text{d}+B_\text{AdS}-\alpha\Delta_B\right) \, .
\end{equation}
For the choice of deviation we take,
\begin{align}\label{eq:IC_Parameterization}
    B_\text{d}&= \Omega_{1}z^4 \cos \left(\gamma_{1} z\right)+\Omega_{2} z^4\tan \left(\gamma_{2} z\right)+\Omega_{3} z^4\sin \left(\gamma_{3} z\right) \nonumber\\
    &+\sum_{i=0}^{5} \beta_{i} z^{i+4}  \, ,
\end{align}
where $\Omega_{\{1,2,3\}}$, $\gamma_{\{1,2,3\}}$ and $\beta_{\{1-5\}}$ are free parameters. One notes that this is precisely the parameterization of the initial data used in~\cite{Rougemont:2021qyk},
\begin{align}
    B_\text{s}\left(z, v_{0}\right)&= \Omega_{1} \cos \left(\gamma_{1} z\right)+\Omega_{2} \tan \left(\gamma_{2} z\right)+\Omega_{3} \sin \left(\gamma_{3} z\right) \nonumber\\
    & +\frac{\alpha}{z^{4}}\left[-\frac{2}{3} \ln \left(1+\frac{z}{v_{0}}\right)+\frac{2 z^{3}}{9 v_{0}^{3}}-\frac{z^{2}}{3 v_{0}^{2}}+\frac{2 z}{3 v_{0}}\right] \nonumber \\
    &+\sum_{i=0}^{5} \beta_{i} z^{i}
    \label{eq:BsInProfileNoronha}
\end{align}
where we have inserted an $\alpha$ into our expression in Eq.~(\ref{eq:init_data}) to match~\cite{Rougemont:2021qyk}.
Furthermore, we select the same parameters as given in table I of~\cite{Rougemont:2021qyk}, reproduced here in table~\ref{tab:IC}. It is important to note that although we do not use the same horizon fixing scheme as~\cite{Rougemont:2021qyk} our time evolution is identical. For more details see appendix~\ref{append:Horizon_Fixing}.

\section{ Out-of-equilibrium speed of sound, entropy and temperature}
\label{sec:SpeedOfSound}
\subsection{Hydrodynamic expectation}
\label{sec:hydro_exp}
Generally, the speed of sound of a relativistic fluid in \emph{global thermal equilibrium} is defined as
\begin{equation}
    c_\text{s}^2 = \left( \frac{\partial P}{\partial \epsilon} \right)_s\,,
\end{equation}
 with the pressure $P$, the energy density $\epsilon$ and the entropy density $s$ (the upright index on $c_\text{s}$ stands for \Say{sound} and the math font index $s$ on the parenthesis stands for entropy density). If we consider an \emph{ideal fluid\footnote{See section~\ref{sec:BjorkenFlow} for the definition of an ideal fluid undergoing Bjorken flow.} boost invariant} evolution then this can be computed by derivatives of the energy-momentum tensor with respect to itself,
 \begin{equation}\label{eq:SoS}
     c_\text{s}^2=-\frac{\partial T^i_i}{\partial T^{0}_0}, \quad T^i_i=P\, ,\quad T^{0}_0=-\epsilon \, ,
\end{equation}
where we recall that for an ideal fluid the energy-momentum tensor with one raised and one lowered index is isotropic. 
Given that the entropy density is a function of only $\tau$, fixing $\tau$ is equivalent to holding entropy density constant. One can then compute the derivative in Eq. (\ref{eq:SoS}) and find $c_s^2=1/3$. This reasoning can be extended to higher orders in the gradient expansion. To begin with, we focus on the \emph{shear stress tensor correction, $\pi_L$} to the $(x_1,x_1)$-component of the energy-momentum tensor which is given by
\begin{equation}
     T^{x_1}_{x_1}= \left(P+\pi_L/2\right)\, ,\quad \pi_L/2=\frac{2 \eta }{3 \tau }-\frac{4 \lambda_1}{9 \tau ^2}+\frac{4 \eta  \tau_\pi}{9 \tau ^2}\, .
\end{equation}
Computing the derivative given in Eq.~(\ref{eq:SoS}) gives\footnote{Both $\epsilon$ and $P$ here depend on $\pi_L$ through solutions to the equations of motion. Once solutions for $\epsilon$ are found $P$ is related to $\epsilon$ via the equation of state $\epsilon=3P$. This behavior is in analogy to the anisotropic equilibrium states generated by a magnetic field, where the pressures and energy density depend on the value of the magnetic field.}
\begin{equation}
    \frac{\partial T^{x_1}_{x_1}}{\partial T^{00}}=\frac{\partial P}{\partial\epsilon } +\frac{1}{2}\frac{\partial\pi_L}{\partial \epsilon}\,.
\end{equation}
where we now make use of the equation of state for a conformal fluid, $P=\epsilon/3$, to find that the first term above gives the zeroth order (in dissipative corrections) to the coefficient $c_\text{s}^2$, referred to as the speed of sound of a conformal fluid, $c_\text{s}^2=1/3$. Now we are left with the computation of the shear component $\pi_L$. To compute its derivative we recall the relation between the hydrodynamic transport coefficients and their dimensionless counterparts~\cite{Romatschke:2017vte} 
\begin{equation}
     \eta= \frac{4 C_\eta\epsilon}{3 T}\, ,\quad   \frac{ \eta \tau_\pi}{\epsilon}=\frac{4}{3}\frac{C_\eta C_\pi}{T^2}\, , \quad    \frac{ \lambda_1}{\epsilon}=\frac{4}{3}\frac{C_\eta C_\pi C_\lambda}{T^2}\, ,
\end{equation}
allowing us to rewrite ($\eta,\tau_\pi,\lambda$) in terms of a relation between temperature and energy. Inserting this into the shear stress tensor component gives
\begin{equation}
   \pi_L=\frac{16C_\eta \epsilon }{9\tau T} -\frac{32 C_\eta  (C_\lambda -1) C_\pi  \epsilon }{27 \tau ^2 T^2}\, .
\end{equation}
We can now directly compute the derivative with respect to energy making use of the relation $T=T_0 \epsilon^{1/4}$ which gives
\begin{equation}
 \frac{\partial \pi_L}{\partial \epsilon} =  \frac{4 C_\eta}{3 \tau  T}-\frac{16 C_\eta (C_\lambda-1) C_\pi}{27 \tau ^2 T^2}\,.
\end{equation}
Altogether, we find the following expressions for the energy derivatives of the energy-momentum components given in terms of the dimensionless transport coefficients valid to second order in the derivative expansion
\begin{align}
   c_\perp^{2,(2)}&=c_\text{s}^2+\frac{2 C_\eta}{3 \tau  T}+\frac{8 C_\eta (1-C_\lambda) C_\pi}{27 \tau ^2 T^2}\, , \label{eq:cperp_hydro}\\
    c_{||}^{2,(2)}&=c_\text{s}^{2}-\frac{4 C_\eta}{3 \tau  T}-\frac{16 C_\eta (1-C_\lambda) C_\pi}{27 \tau ^2 T^2}\, ,\label{eq:cpara_hydro}
\end{align}
where the superscript ``$(2)$'' indicates that these are the second order corrected speeds of sound. This second as well as the first order correction to the conformal speed of sound stem from the fact that the \emph{viscous} plasma is expanding in longitudinal direction. Hence, the medium on which perturbations are propagating is changing its pressure over time, and acquires different pressures in longitudinal and transverse directions. These changes in the pressures of the plasma lead to a change in the propagation speed of longitudinal waves, such as the sound waves.

\subsection{Holographic out-of-equilibrium calculation}\label{sec:out_of_eq}
\begin{figure}
    \centering
    \includegraphics[width=0.49\textwidth]{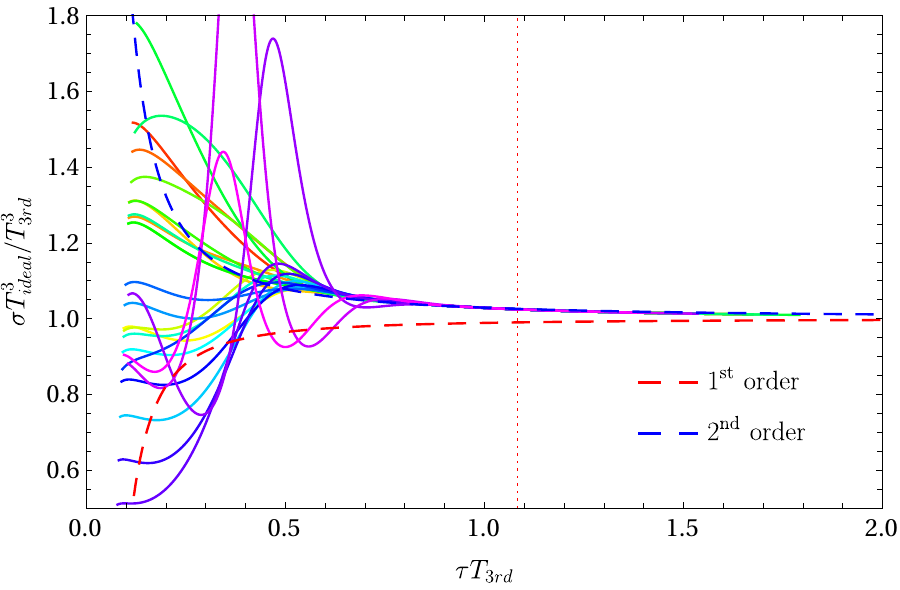}\hfill
    \includegraphics[width=0.49\textwidth]{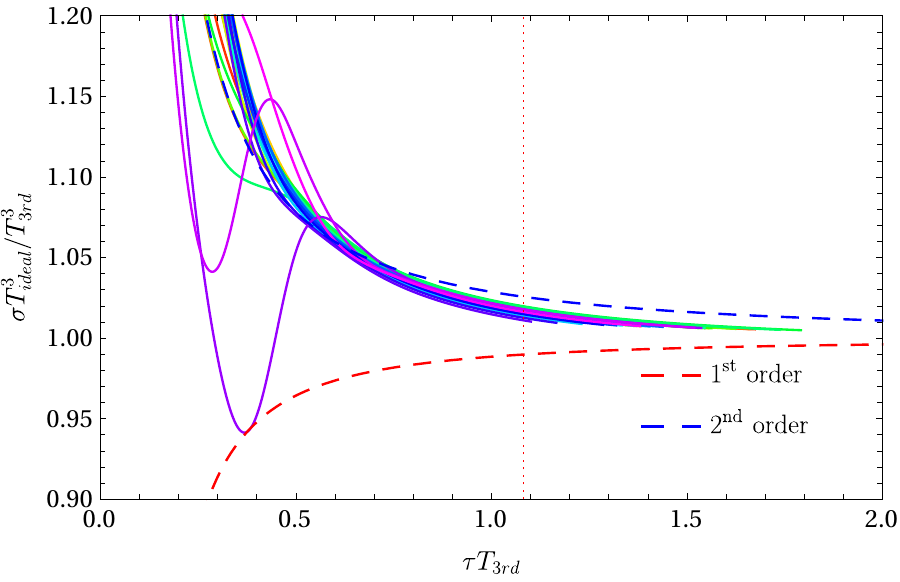}
    \caption{\textit{Scaled Entropy Density:} 
    The scaled entropy $\sigma$ defined in equation~(\ref{eq:entropy}) (see~\cite{Rougemont:2021qyk}) is displayed for all the initial conditions considered in this work.
    The red and blue dashed lines indicate the first and second order hydrodynamic expansion from~\cite{Rougemont:2021qyk}. \Left Dimensionless entropy normalized with the dynamical temperature, $T$, computed from the Stefan-Boltzmann law (see Eq. (\ref{eq:dyn_T})). This entropy measure agrees from early times onward with the second order hydrodynamic prediction (blue dashed curve)~\cite{Rougemont:2021qyk}. \Right Dimensionless entropy normalized with the temperature, $T_{3rd}$, computed from the hydrodynamic equations. 
    \label{fig:Scaled_Entropy}}
\end{figure}

Given our hydrodynamic expectation derived above, our goal now is to compute
\begin{equation}\label{eq:outOfEqCs}
   c^2_{\perp} =- \frac{\partial\vev{T^{x_1}_{x_1}}}{\partial \vev{T^0_0}}\, , \qquad c^2_{||}=- \frac{\partial\vev{T^\xi_\xi}}{\partial \vev{T^0_0}}\, ,
\end{equation}
in the holographic model. We have tested four ways to do this using only the energy-momentum tensor which 
is given in terms of our numerical data:  
\begin{enumerate}
    \item The na\"{i}ve way, making use of a chain rule, varying the energy and pressure separately as functions of time. 
    \item Direct variation of the pressure and energy in the holographic model on \emph{fixed entropy slices}. 
    \item Direct variation of the pressure and energy in the holographic model on \emph{fixed apparent horizon area slices}.
    \item Direct variation of the pressure and energy in the holographic model on \emph{fixed time slices}. 
\end{enumerate}
Each of these methods suffers from its own deficiencies.  
The simplest of these methods of calculating the speed of sound is the first, based on the chain rule
\begin{equation}
    c_\text{s}^2 \sim \frac{\partial P}{\partial \tau} \left( \frac{\partial \epsilon}{\partial \tau} \right)^{-1}\, .
    \label{eq:naiveSpeedOfSound}
\end{equation}
However, if our goal is to keep the entropy fixed as is done in the equilibrium calculation then this is clearly not fully correct since along $\tau$ the entropy changes. At least in equilibrium the dual field theory entropy is given by the area of the black hole event horizon~\cite{tHooft:1993dmi,Susskind:1994vu}. However, motivated by advances in the understanding of fluid dynamics in the AdS/CFT duality, which gave rise to the fluid/gravity correspondence~\cite{Bhattacharyya:2008jc,Bhattacharyya:2008xc}, numerous authors have considered how to define out-of-equilibrium entropy. Many of these notions are based on trapped surfaces~\cite{Booth:2010kr,Booth:2011qy} including the most popular workhorse of the community defined by the area of the outer-most trapped surface, or apparent horizon, whose area is proposed as the relevant one dual to the field theory entropy~\cite{Engelhardt:2017aux}. In our coordinates the field theory entropy associated with the apparent horizon can be computed from the ratio of the apparent horizon area, $A_\text{AH}$
\begin{align}
    S(\tau)&= \frac{1}{4G}A_\text{AH} =\frac{1}{4G}\int \dd^3x \sqrt{-g} \nonumber \\
    &=\frac{1}{4G} S(z_\text{AH},\tau)^3\int \dd x\dd y\dd \xi \, ,
\end{align}
and the field theory area $\mathcal{A}$ (with field theory metric $\gamma$)
\begin{equation}
    \mathcal{A}=\int \dd^3x \sqrt{-\gamma}  =\tau \int\dd x\dd y\dd \xi \, ,
\end{equation}
and is given by
\begin{equation}
     s(\tau)= \frac{1}{4G}\frac{A_\text{AH}}{\mathcal{A}}= \frac{S(z_\text{AH},\tau)^3}{4G\tau}\,.
\end{equation}
While we can work directly with $s(\tau)$ is useful to construct a dimensionless entropy density\footnote{
    We call it $\sigma$ because it is like a dimensionless entropy density and $\sigma$ is the lower letter s in
    the Greek alphabet.} $\sigma$ defined as~\cite{Rougemont:2021qyk} 
     \begin{equation}
        \sigma(\tau) \equiv \frac{s(\tau)}{2 \pi^4 T^3_\text{ideal}(\tau)} = \frac{A_\text{AH}(\tau)}{\pi^3 \Lambda^2 \mathcal{A}}
         = \frac{|S(z_\text{AH}, \tau)|^3}{\pi^3  \Lambda^2}\,.
         \label{eq:entropy}
    \end{equation}
This can be seen in Fig.~\ref{fig:Scaled_Entropy} where we have plotted the dimensionless entropy density scaled with the true out-of-equilibrium temperature $T$, rather than that of an ideal boost invariant fluid\footnote{When normalized to the temperature as obtained from the hydrodynamic expansion the entropy density never converges to a single curve. Instead the curves from all of the evaluations come to a band. When normalized to the temperature as obtained from the Stefan-Boltzmann law the entropy density from each curve collapses to a single curve at approximately $\tau T=1$. 
There seems to be a maximal possible entropy density around $\sigma \approx 0.8$ for early times. The curves close to this value at early times seem to not produce any entropy until they are close to the attractor at around $tT\approx 1.2$.
Notice that the vertical red line, which indicates the attractor behavior for $f$, $\Delta p/\epsilon$ and $c_{\perp,\para}^2$ only occurs when the entropies have already long converged to one curve, indicating the hydrodynamic behavior starts earlier for the entropy than for the other quantities. We note that our data/analysis agrees with Jakub Jankowski (private communication).  }.
For a static, planar, Schwarzschild black brane in $AdS_{4+1}$ one finds a Stefan-Boltzmann like relation between the energy density $\epsilon$ and the temperature $T$ as $\epsilon=\sigma_\text{SB}T^4$. The Stefan-Boltzmann constant $\sigma_\text{SB}$ in this case is given by~\footnote{For more information see~\cite{Natsuume:2014sfa}. } $\sigma_\text{SB}=\frac{3 \pi ^3 L^3}{16 G}$ in terms of gravitational data or $\sigma_\text{SB}=\frac{3 \pi ^2 N_c^2}{8}$ in terms of field theory data. We make use of this relation to define $T$ out of equilibrium and in our notation in appendix~\ref{appendix:Numerics} is given as 
\begin{equation}
T=a_4^{1/4}/\pi \, . \label{eq:dyn_T}
\end{equation}

Although the entropy density is time dependent (as seen in Fig.~\ref{fig:Scaled_Entropy}) we see that for late times the entropy only varies slowly, so for late times the above notion of the speed of sound in Eq. (\ref{eq:naiveSpeedOfSound}) should at least converge to the correct values expected from Eq. (\ref{eq:cperp_hydro}) and Eq. (\ref{eq:cpara_hydro}). However early during our time evolution the entropy changes rapidly in time. We may then suspect that these method should not reliably describe the true speed of sound. The second and third methods of computing the speed of sound in our list were generated with this mind. These methods make use of direct differences in the pressure and energy at fixed values of either the field theory entropy $s(\tau)$ or the apparent horizon area $\sigma$ (where one notes $\sigma$ is indeed the horizon area modulo a numerical factor). 

To facilitate the calculation of the speed of sound using method 2 or 3 in our list we begin by noting that both $\tau$ and $\sigma (s)$ are monotonic functions, hence we can switch from using $\tau$ to $\sigma (s)$ as our temporal coordinate. From our time evolution we have $\epsilon(\tau)$, $P_\perp(\tau)=T^1_1=T^2_2$, $P_{\para}(\tau)=T^\xi_\xi$ as well as $\sigma(\tau)$ and hence we can construct $\epsilon(\sigma)$ and $P_{\perp,\para}(\sigma)$. To facilitate the derivative we now compute an array of $N$ curves with initial energies in the interval $(\epsilon_0-\Delta \epsilon, \epsilon_0 + \Delta \epsilon)$ with the variation in energy $\Delta \epsilon$ small. 
  
With our family curves computed we denote by $\epsilon(\sigma_i, \epsilon_0 -\Delta \epsilon + j \delta\epsilon)$ the energy density at entropy $\sigma_i = \sigma(\tau_i)$ (recall $\tau$ is a member of an evenly spaced grid constructed during the time evolution) whose initial energy density was $\epsilon_0 -\Delta \epsilon + j \delta\epsilon$ with $0<\delta \epsilon \leq \Delta \epsilon$. Since the evolution of the entropy depends also on the initial energy not every curve has the same dimensionless entropy $\sigma_i$. We then interpolate each curve in the family to obtain smooth functions $\epsilon(\sigma, \epsilon_0 -\Delta \epsilon + j \delta\epsilon)$ of the dimensionless entropy. We are now free to construct a uniform grid of dimensionless entropy $\sigma \in [\sigma_\text{initial},\sigma_\text{final}]$ on which we can evaluate our smooth functions. This ensures that now at each instant of entropy, $\sigma_i$, we have $N$ values for $\epsilon$, one value for each member of the family.

For each instant in entropy $\sigma_i$, we may use the family of curves $\epsilon_j(\sigma_i) := \epsilon(\sigma_i, \epsilon_0 -\Delta \epsilon + j\delta \epsilon)$ to construct a finite difference representation of the derivative\footnote{We have also used Mathematica's \textbf{DerivativeFilter} function~\cite{Mathematica}, which uses a spline interpolation to represent the numeric derivative to check that indeed a finite difference is sufficient. To use this function we found that 9 curves accurately represent the derivative.}. given the pressures $P_\perp(\sigma_i, \epsilon_0 -\Delta\epsilon + j \delta \epsilon)$ and $P_{\para}(\sigma_i, \epsilon_0 -\Delta\epsilon + j \delta \epsilon)$. We use the centered differences rule
\begin{equation}\label{eq:central_diff}
      \frac{\dd P_{ j}(\sigma_i)}{\dd \epsilon_j(\sigma_i)}=\frac{\left( P_{j+1}(\sigma_i) - P_{j-1}(\sigma_i) \right)}{\left( \epsilon_{j+1}(\sigma_i) - \epsilon_{j-1}(\sigma_i) \right)} \, , 
\end{equation}
which we emphasize, by definition, is at constant $\sigma$. The final step of the procedure is to transform back to $\tau$ as time variable. This requires that we invert $\sigma(\tau)$ to get $\tau(\sigma)$. We do this by interpolating $(\sigma_i, \tau_i)$ noting that this must be done for each curve separately. We can use this procedure to construct the speed of sound for each curve $j$ at time $\tau_i$, 
\begin{equation}\label{eq:Speed_of_sound_NONEQ}
        c_{\perp,\, j}(\tau_i)^2=\frac{\dd P_{\perp, j}(\tau_i)}{\dd \epsilon_j(\tau_i)},\quad c_{\para,\, j}(\tau_i)^2=\frac{\dd P_{\para, j}(\tau_i)}{\dd \epsilon_j(\tau_i)}\,.
\end{equation}

Finally, this can be compare to method 4, direct variation of the energy. This is done similar to method 2 and 3 where we now compute an array of curves with initial energies in the interval $(\epsilon_0-\Delta \epsilon, \epsilon_0 + \Delta \epsilon)$. However rather then work at fixed slices of the entropy or horizon area we instead hold $\tau$ fixed and directly compute the speeds of sound via a central difference as in Eq. (\ref{eq:central_diff}). 

\begin{figure}[t]
    \centering
     \includegraphics[width=0.49\textwidth]{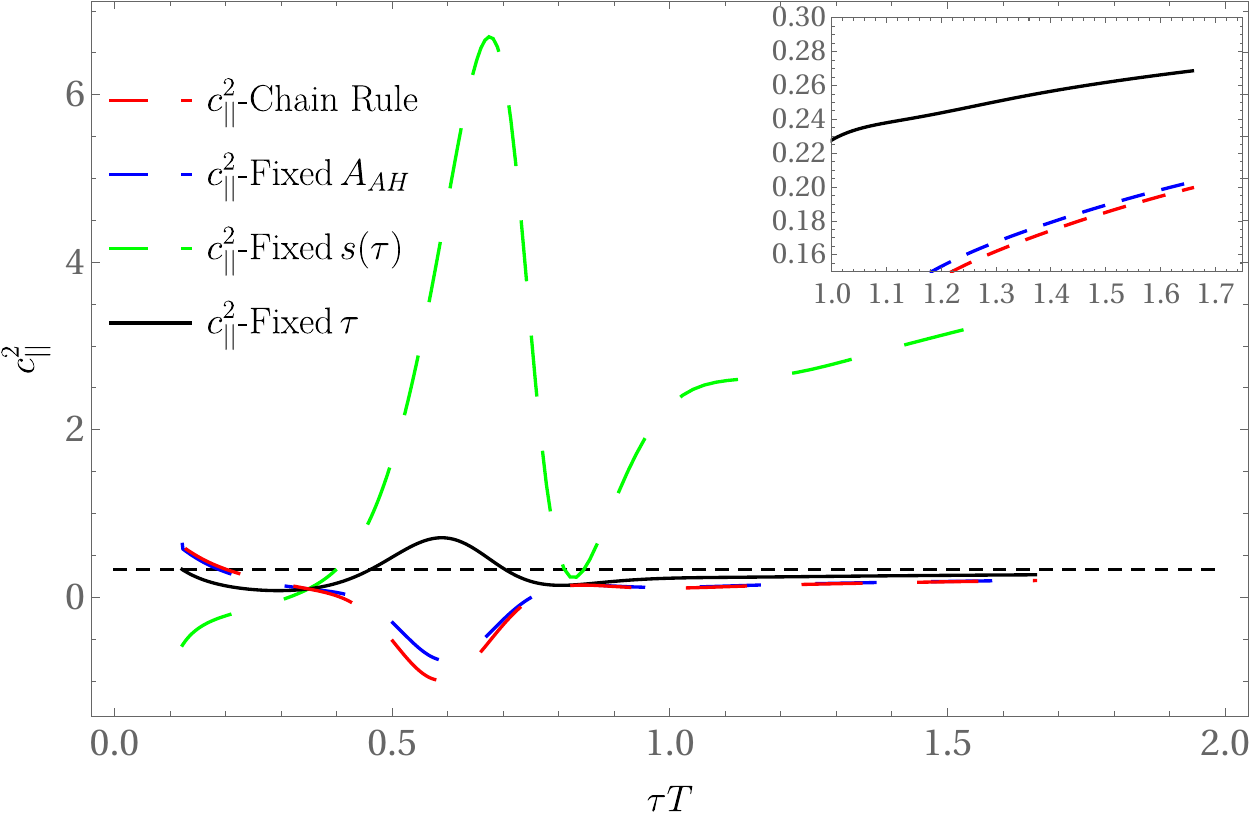}\hfill
    \includegraphics[width=0.49\textwidth]{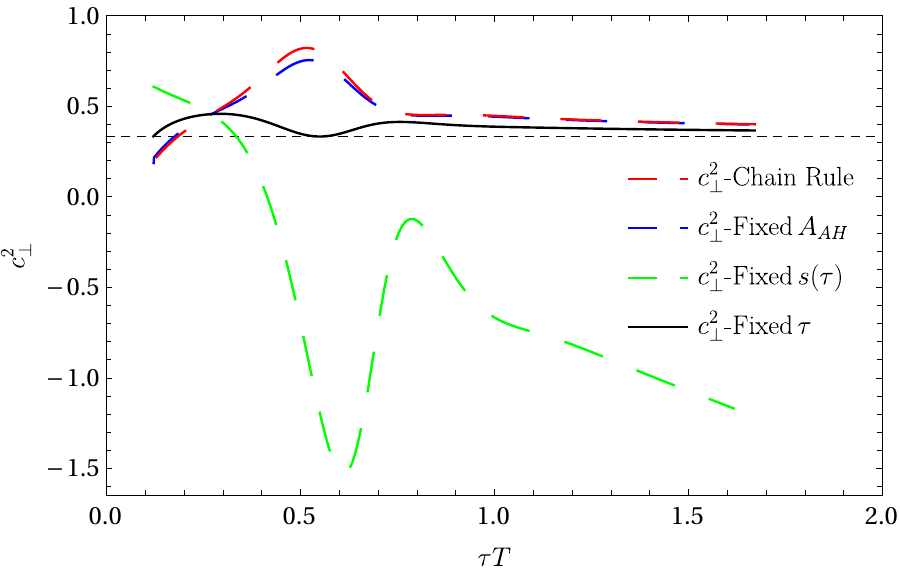}
    \caption{{\it Comparison of Methods:} The speed of sound as computed via 4 distinct methods is displayed. \Left $c_{\para}^2$. \Right $c_\perp^2$. In both images the red curve displays the chain rule method, the blue dashed curve displays holding the apparent horizon area fixed, the green curve displays holding the field theory entropy fixed, and the black line displays holding $\tau$ fixed. 
    \label{fig:Methods_Comparison}}
\end{figure}

\begin{figure}
    \includegraphics[width=0.49\textwidth]{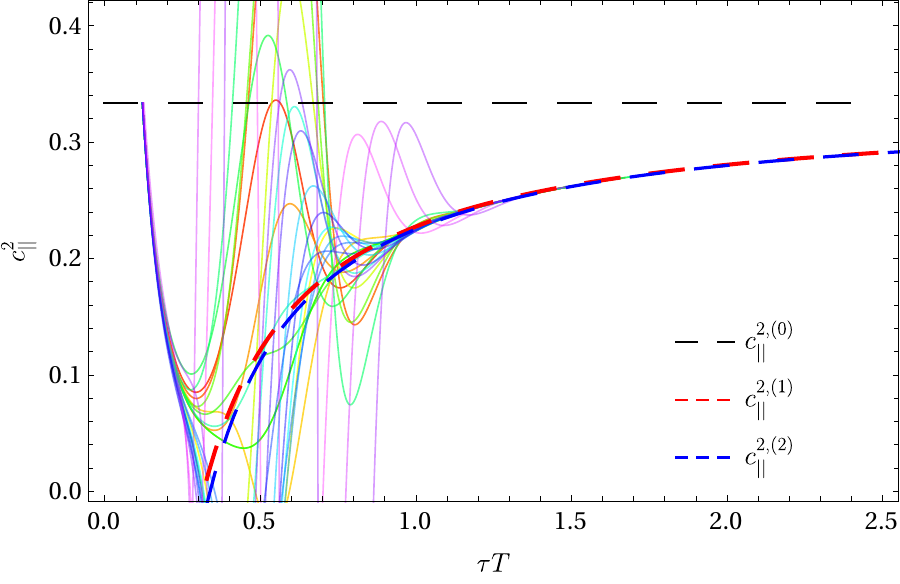}\hfill \includegraphics[width=0.49\textwidth]{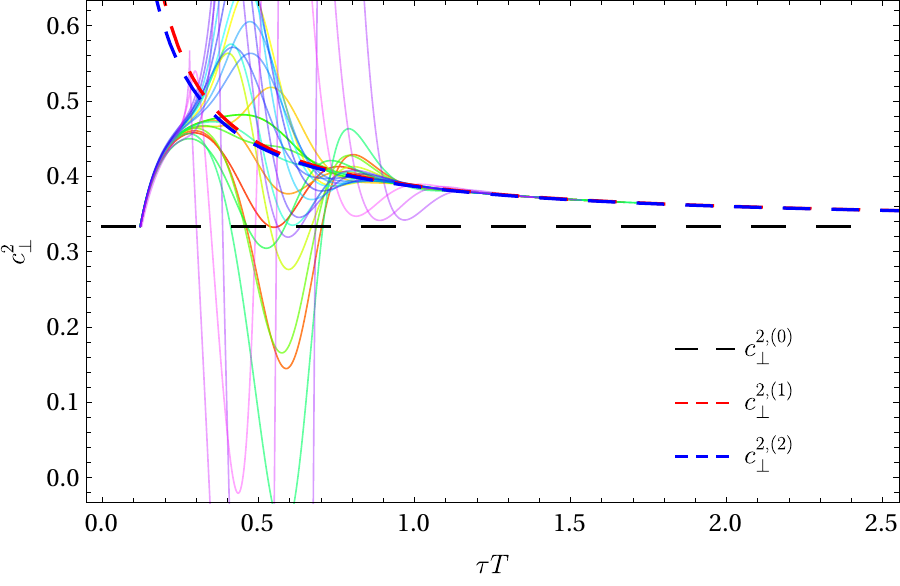}
    \caption{\textit{Hydrodynamic Comparison:} 
    The images display the speed of sound as computed via the full evolution scheme. \Left $c_{\para}^2$. \Right $c_\perp^2$. The hydrodynamic expectation of the thermodynamic derivatives are shown as (Red - 1st order in the derivative expansion, Blue - 2nd order in the derivative expansion) dashed lines. The conformal speed of sound in the system is displayed as a dashed black line $c_\text{s}^2=1/3$.
    \label{fig:hydro_comp}
    }
\end{figure}

Having described all four methods we can compare the result of these as is shown in Fig.~\ref{fig:Methods_Comparison}. 
Shockingly the worst faring of the four methods is the one at fixed field theory entropy $s(\tau)$. This method can be seen growing/decreasing over time past $\tau T>1$ linearly in $\tau T$. While the remaining three methods roughly behave in a similar manner beyond $\tau T>1$. Shown in the inset graphic of the left image of Fig.~\ref{fig:Methods_Comparison} is a closer look at the region $1<\tau T<2$ of $c_\perp^2$. Here it can be seen that although all three behave in a similar manner there is an offset between each of these three curves. Of these three we will focus on method 4, direct calculation at fixed $\tau$ going forward. The results for the speed of sound using method 4 are displayed in Fig.~\ref{fig:hydro_comp}. In order to generate the plots, the time evolution has been performed for all initial conditions from~\cite{Rougemont:2021qyk} as displayed in table~\ref{tab:IC} with $\Delta\epsilon=0.0075$.
One can see that the proposed speeds of sound do not fully approach the conformal value of $c^2=1/3$, rather $c_\perp^2$ stays slightly above and $c_{\para}^2$ stays slightly below. One can note that the variation of the speed of sound as computed with this method can vary wildly. 
The curves for some initial conditions even display superluminal speeds and others display instabilities, indicating that these initial conditions should be discarded. Interestingly, we will see that this seems to be related to the violation of the weak energy condition in those cases.
Furthermore, with all methods, the out-of-equilibrium speed of sound in the transverse direction is non-trivial, as 
expected due to the anisotropy in the system as a result of the non-trivial shear stress tensor.

In order to understand the merit of the calculation one should compare the holographic calculation with the hydrodynamic expectation as computed in section~\ref{sec:hydro_exp}. This comparison is displayed in Fig.~\ref{fig:hydro_comp} as dashed lines. Here one can see clearly that despite the rapid fluctuations of the speed of sound as computed in the holographic method they quickly approach the hydrodynamic expectation in Eq. (\ref{eq:cpara_hydro}) and Eq. (\ref{eq:cperp_hydro}).  

\begin{figure}
    \centering
    \includegraphics[width=0.4\textwidth]{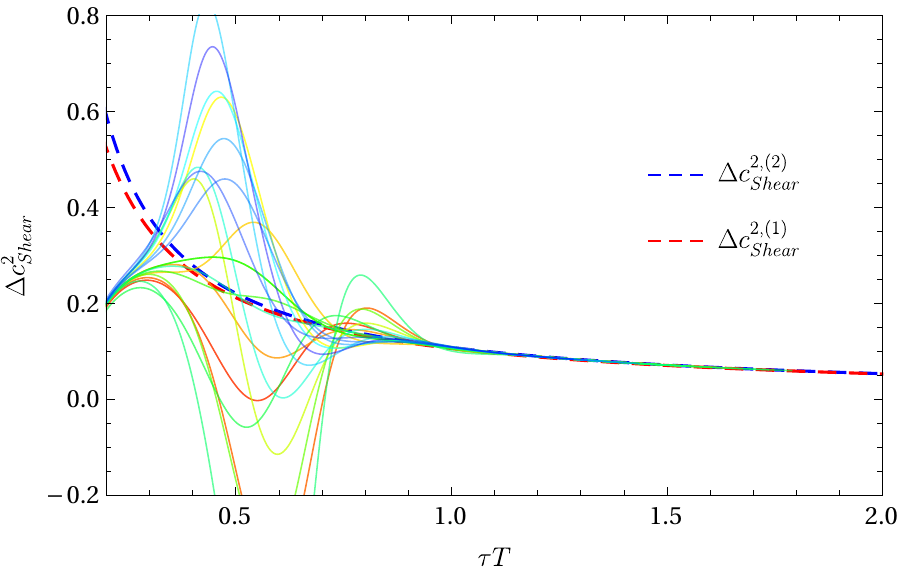}
    \caption{\textit{Shear component of the speed of sound:} 
    The image displays the derivative of the shear stress tensor with respect to the energy. The lines indicate isolating the shear component of the energy-momentum tensor from the full holographic evolution as done in Eq.~(\ref{eq:isolate}). The hydrodynamic expectation of the derivative of the shear stress tensor with respect to the energy is shown as dashed lines (Red - 1st order in the derivative expansion, Blue - 2nd order in the derivative expansion).
    \label{fig:shear_contributions_isolated}
    }
\end{figure}
It is interesting to note that the ``out of equilibrium" component of the speed of sound in both the transverse and longitudinal direction in the hydrodynamic calculation is due solely to the contribution of the shear stress tensor (as can be see in Eq.~(\ref{eq:cpara_hydro}) and Eq.~(\ref{eq:cperp_hydro})). One can isolate the ideal and shear stress component of the speed sound via
\begin{equation}\label{eq:isolate}
  c_\text{s}^2=\frac{1}{3}\left(2c_{\perp}^2+c_{\para}^2\right)\, ,\quad \Delta c_\text{shear}^2\equiv\frac{\partial \pi_{L}}{\partial \epsilon}=\frac{2}{3}(c_\perp^2-c_{\para}^2) \, .
\end{equation}
Our numerical data gives $c_s^2=1/3$ for the entire time evolution using Eq.~(\ref{eq:isolate}) while the shear component is displayed in Fig.~\ref{fig:shear_contributions_isolated}. There one sees that the shear components quickly approach the 2nd order result expected from applying Eq.~(\ref{eq:isolate}) to Eq.~(\ref{eq:cpara_hydro}) and Eq.~(\ref{eq:cperp_hydro}). 

\subsection{Thermodynamic consistency}\label{sec:thermo}
A fundamental idea in the study of hydrodynamics is the notion of local thermal equilibrium. 

\paragraph{Local versus global thermal equilibrium} Let us define \emph{local thermal equilibrium}. The process of reaching local thermal equilibrium will be referred to as \emph{local thermalization}. 
\begin{center}
    \parbox{0.4\textwidth}{
        Local thermal equilibrium is reached at a spacetime point $x$ within a fluid, if and only if the thermal quantities energy density $\epsilon(x)$, temperature $T(x)$, pressure $P(x)$, entropy density $s(x)$ can be defined within a fluid volume element located at $x$, \emph{and if} these quantities take their \emph{local equilibrium values} at that location $x$. This definition of local thermalization implies that the thermodynamic relation $\epsilon(x)+P(x)=s(x)\, T(x)$ is satisfied locally at $x$. This may also be referred to as \emph{local thermodynamic consistency}.
    } 
\end{center}
This definition is in line with the principle that in global thermal equilibrium all observables take their equilibrium values~\cite{Pathria:1996hda,DAlessio:2015qtq}. 
In analogy to that, we define local thermal equilibrium as the state in which all local observables reach local equilibrium values. 
Note, that any neighboring fluid element at $x_1=x+\Delta x$ can have vastly different values of $\epsilon(x_1)\neq \epsilon(x),\, T(x_1)\neq T(x),\,  ...$, allowing for large spatial and time gradients. In other words, local thermal equilibrium could be reached, while the system displays large gradients, indicating that it is far from \emph{global thermal equilibrium}. 

Global thermal equilibrium is reached when the \emph{ergodic hypothesis} is satisfied, i.e.\ the system had sufficient time to explore all of the phase space accessible to it under the given macroscopic constraints~\cite{Pathria:1996hda,DAlessio:2015qtq}.\footnote{Several example systems have been rigorously proven to satisfy the ergodic hypothesis~\cite{Sinai_1970,Bunimovich:1979,Simanyi:2004}.} 
In that case, the standard concepts of statistical mechanics apply and the time-averaged values of all observables are equal to their ensemble-averaged values. This allows the standard technique of considering multiple fictitious copies of a system, a thermodynamic ensemble, and computing ensemble averages~\cite{Pathria:1996hda} instead of long-time averages which are often more difficult to compute. By definition, the time it takes a system to reach ergodicity is long compared to all scales in the system.\footnote{
Alternately, thermodynamic equilibrium may be reached much faster, by the principle of \emph{typical configurations}~\cite{DAlessio:2015qtq}. According to that principle, almost all accessible microscopic configurations  the system can assume are macroscopically equivalent, producing the same values for all observables. These configurations are called \emph{typical}. There exist only few non-typical configurations, which relax quickly to a typical configuration. This principle of typical configurations dominating is adopted in the Eigenstate Thermalization Hypothesis (ETH)~\cite{DAlessio:2015qtq}, 
however, has less rigorous support than the ergodic hypothesis~\cite{DAlessio:2015qtq}. 
}

\begin{figure}[t]
    \includegraphics[width=0.4\textwidth]{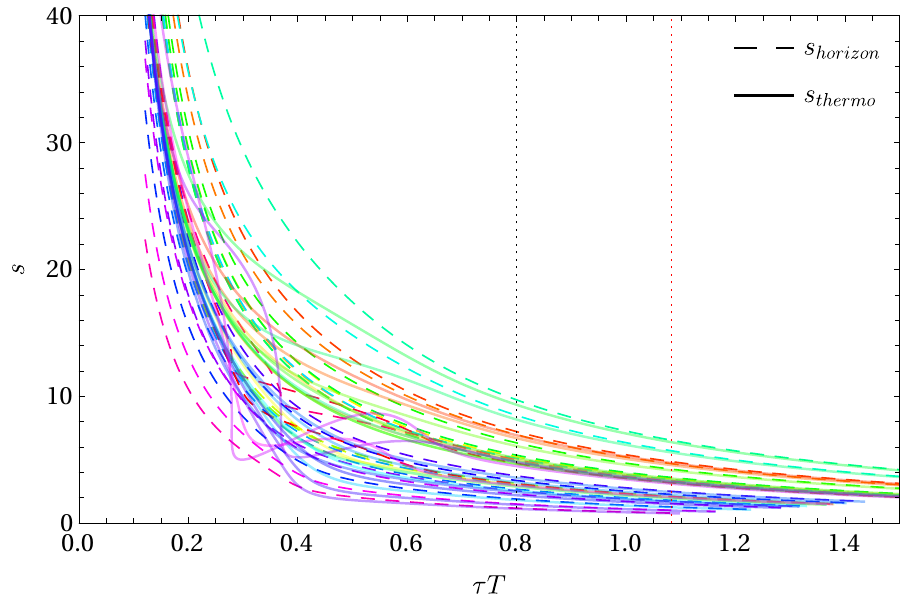}
    \caption{\textit{Thermodynamic Consistency:} A comparison of the entropy density as obtained from the thermodynamic Euler relation with the entropy density obtained from the apparent horizon. The thermodynamic estimated was computed by isolating the \emph{isotropic} component of the energy-momentum tensor $P=\frac{1}{3}(2T^1_1+T^\xi_\xi)$. The dashed lines indicate $s$ as obtained from the apparent horizon while the solid lines indicate the $s$ as obtained from the Euler relation. Clear deviations from between the solid and dashed curves can be seen for times $\tau T\lesssim 1$.  While for $\tau T\gtrsim 1$ the dashed and solid curves differ but both follow the same falloff. For larger times the curves continuously come closer to agreement. 
    }
    \label{fig:Thermo_AH}
\end{figure}
\begin{figure}
    \centering
   \includegraphics[width=0.49\textwidth]{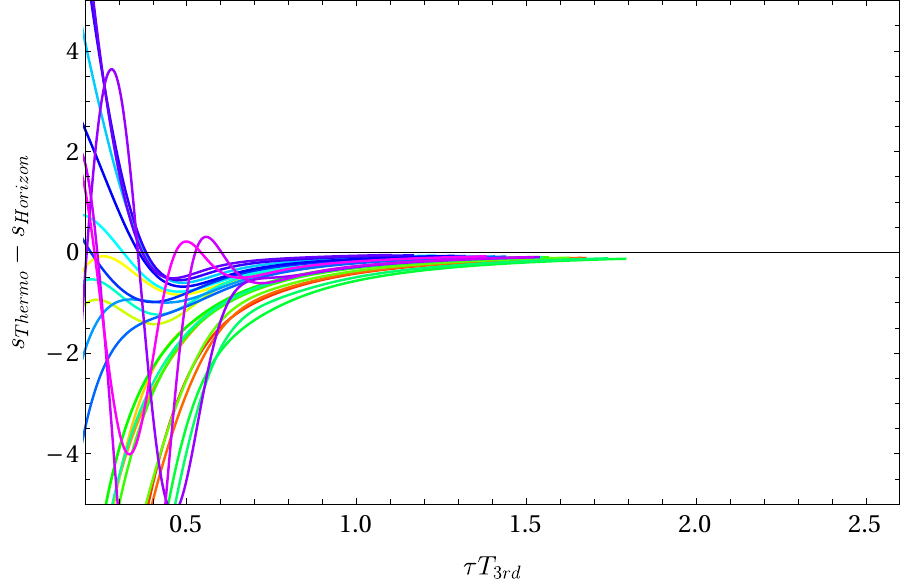}\hfill \includegraphics[width=0.49\textwidth]{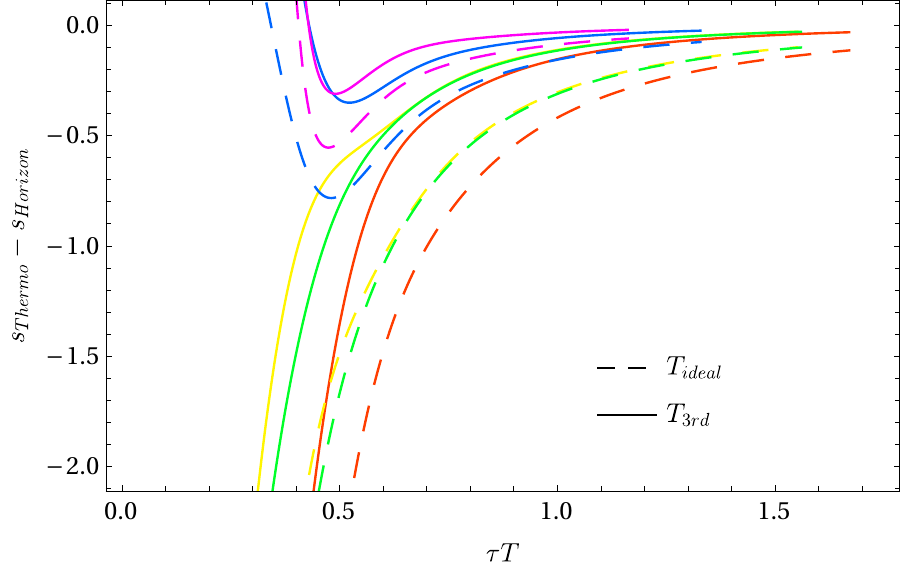}
    \caption{\textit{Thermodynamic Consistency - differences:} The difference between the entropy density as obtained from the thermodynamic Euler relation ($s_\text{thermo}$) with the entropy density obtained from the apparent horizon ($s_\text{horizon}$). \Left The difference is computed with $T=T_\text{ideal}=\tilde{\Lambda}^{2/3}/\tau^{1/3}$. \Right The difference is computed with $T=T_\text{3rd}$ displayed with solid lines. In the right image the difference is also displayed for $T=T_\text{ideal}$ as dashed lines. Notice that the solid line is always closer to zero then the dashed line of the corresponding color. This indicates that using $T_\text{3rd}$ in the equation for the thermodynamic entropy density is closer to value of the entropy density as computed from the apparent horizon. In both cases the Euler relation was computed with $P_{\para}$. 
    \label{fig:Thermo_AH_diff} }
\end{figure}

The standard definition of the speed of sound is valid only in global thermal equilibrium. Clearly we are not in a state of global thermal equilibrium but perhaps we are in local equilibrium. To measure whether or not this is the case we can use the local thermodynamic Euler relation, as defined above, now with $x\to\tau$ as this is the only coordinate dependence consistent with Bjorken symmetry 
\begin{equation}
    \epsilon(\tau)+P(\tau)= s(\tau)T(\tau) \, 
\end{equation}
as a reference. Using the Euler relation one obtains the entropy density as $s=(\epsilon+P)/T$ which can be compared directly with the entropy as computed from the apparent horizon. The results of the calculation are displayed in Fig.~\ref{fig:Thermo_AH}. The dashed lines in the figure represent the entropy density as obtained from the apparent horizon while the solid lines represent the entropy density obtained from the Euler relation. Clearly the Euler relation, a thermodynamic equation, will not be a valid equation throughout the full evolution. Indeed there are large deviations between the entropy density computed from the Euler relation and that from the apparent horizon for $\tau T\lesssim 1$.
\begin{figure}[t]
    \centering
    \includegraphics[width=0.49\textwidth]{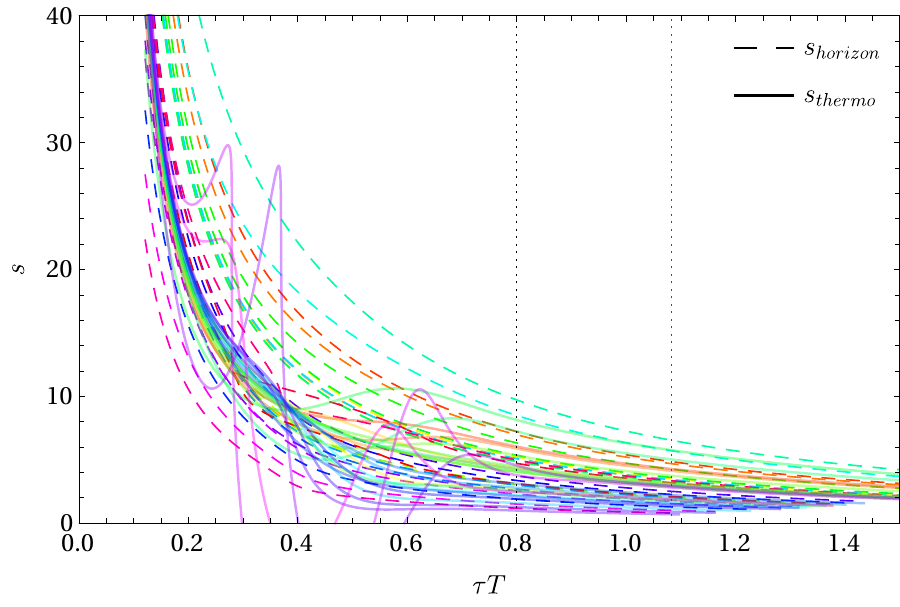}\hfill \includegraphics[width=0.49\textwidth]{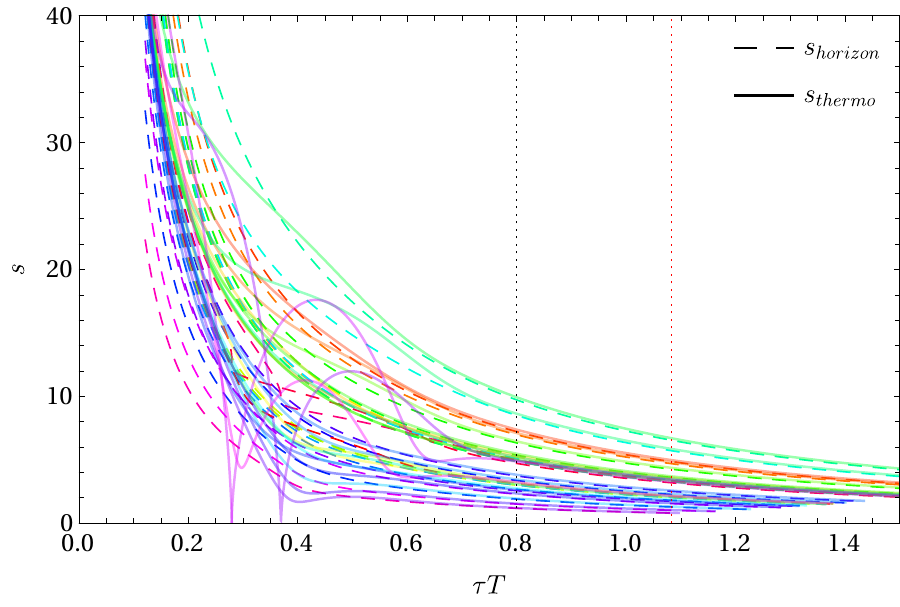}
    \caption{\textit{Thermodynamic Consistency:} A comparison of the entropy density as obtained from the thermodynamic Euler relation with the entropy density obtained from the apparent horizon. \textit{Left:} The Euler relation computed using $P_{\para}=T^\xi_\xi$. \textit{Right:} The Euler relation computed using $P_\perp=T^i_i$ for $i=1,2$. The dashed lines indicate $s$ as obtained from the apparent horizon while the solid lines indicate the $s$ as obtained from the Euler relation. Clear deviations from between the solid and dashed curves can be seen for times $\tau T\lesssim 1$.  While for $\tau T\gtrsim 1$ the dashed and solid curves differ but both follow the same falloff. For larger times the curves continuously come closer to agreement. 
    }
    \label{fig:Thermo_AH_PL_PT}
\end{figure}
However for $\tau T\gtrsim 1$ the entropy as computed via the Euler relation quickly begins approaching the entropy density computed from the apparent horizon. This is further displayed in Fig.~\ref{fig:Thermo_AH_diff} where we have displayed the difference between the thermodynamic entropy density $s_\text{thermo}=(\epsilon+P)/T$ and the entropy density as computed from the apparent horizon $s_\text{horizon}$. On the left this is computed using the ideal temperature as obtained in $T(\tau)= \tilde{\Lambda}^{2/3}/\tau^{1/3}$. 
From Fig.~\ref{fig:Thermo_AH_PL_PT}, we see that the pressure that satisfies the Euler relation the earliest is the \emph{isotropic} pressure $P$, followed by the transverse $P_\perp$, and the worst is the longitudinal $P_{\para}$. This may suggest that the isotropic pressure $P$ is a candidate for an out-of-equilibrium generating functional in this case. 

As in Fig.~\ref{fig:Thermo_AH}, one sees in the left image of Fig.~\ref{fig:Thermo_AH_diff}  after a time of $\tau T_{\text{3rd}}\gtrsim 1$ all of the curves begin a universal trajectory, asymptotically approaching zero. The image on the right in Fig.~\ref{fig:Thermo_AH_diff} displays that this approach to agreement between thermodynamic and horizon based entropy densities is faster when taking into account further terms of the hydrodynamic derivative expansion. This can be seen by noticing that the solid lines (representing the difference computed with $T=T_\text{3rd}$) is closer to the axis for all curves displayed then corresponding dashed line (representing the difference computed with $T=T_\text{ideal}$). 

In summary what we have found in the series of images displayed in Fig.~\ref{fig:Thermo_AH}-\ref{fig:Thermo_AH_diff} is: 
\begin{enumerate}
    \item  $s_\text{horizon}$ agrees with hydrodynamic expectations after a time scale of approximately $\tau T\gtrsim 0.8$.
    \item  $s_\text{horizon}$ agrees reasonably well with thermodynamic expectations after a time scale of approximately $\tau T\gtrsim 1$. 
\end{enumerate}
Taken together we can expect a thermodynamic definition of the speed of sound to agree hydrodynamic expectations after a similar amount of time has passed in the evolution of the system. This is exactly what is seen in Fig.~\ref{fig:hydro_comp}. Furthermore it is now interesting to compare this time scale to what has already been observed in the literature. In Fig.~\ref{figs:deltaPAndc} on the left hand side we show the pressure anisotropy divided by the energy density as a function of $\tau T$. This has been studied for instance  in~\cite{spalinskiHydrodynamicAttractorYangMills2018} in the context of attractor solutions in hydrodynamics. The blue dotted line in the left image shows the analytic form for the attractor obtained in~\cite{spalinskiHydrodynamicAttractorYangMills2018}. The red vertical line in the figures indicates an approximate time when the attractor behavior sets in. Here we see this is exactly the same time at which the attractor behavior of the speeds of sound $c_{\perp,\para}^2$ begin. Furthermore it is exactly the time scale when $s_\text{horizon}$ agrees with hydrodynamic expectations and reasonably agrees with thermodynamic expectations.

\begin{figure}
    \centering
    \includegraphics[width=0.4\textwidth]{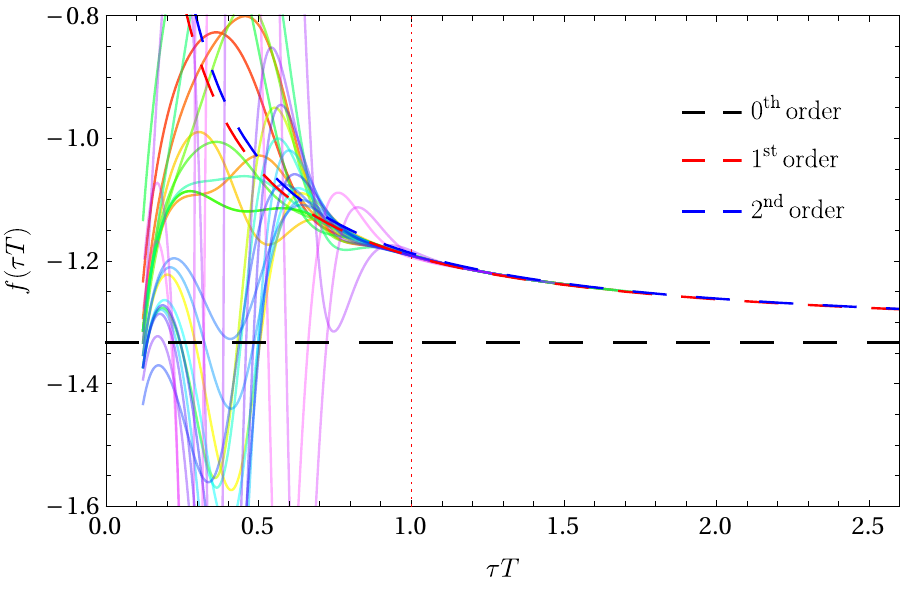}
    \caption{The function $f$ as a function of $\tau T$ together with 0th, 1st and 2nd order expansions
    given in equation~(\ref{eq:BjorkenEpsilon2}) with $T\to T_\text{3rd}$ and $C_\eta = 1/(4\pi)$, $C_\pi = 0.21$, $C_\lambda = 0.77$.}
    \label{fig:fplot}
\end{figure}

\subsection{Borel resummation}
\begin{figure}
    \centering
    \includegraphics[width=0.4\textwidth]{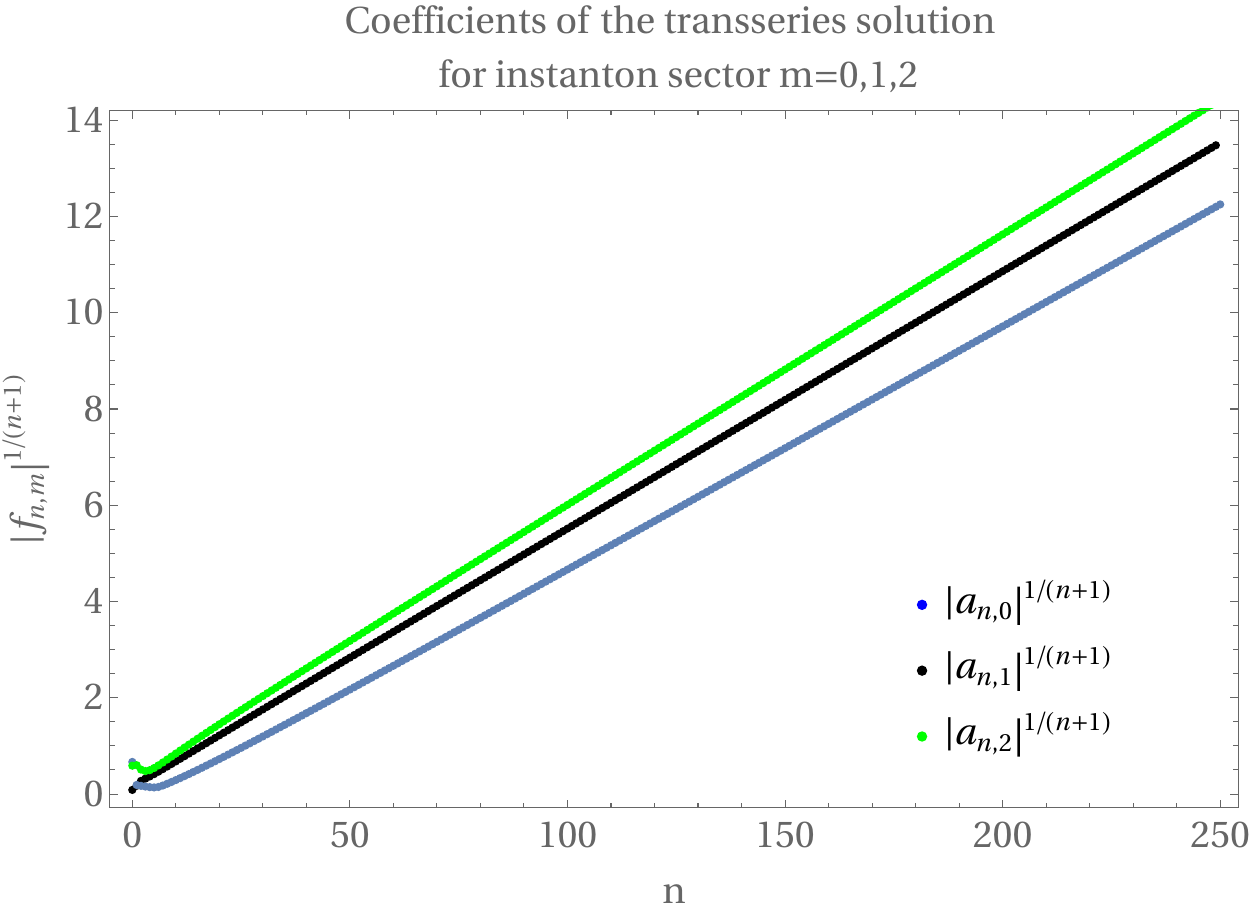}
    \caption{\textit{Coefficients of the Transsseries:} The coefficients $a_{n,m}$ of the transseries in Eq.~(\ref{eq:trans}) are displayed for the first three instanton sectors. One can see they each have a distinct slope which can be checked to coincide with the distance to the nearest singularity. 
    \label{fig:Divergent}}
\end{figure}

The formulation of hydrodynamics with which we have worked with up until this point has been the standard Landau-Lifschitz formulation. However, working to finite order in the corrections (for instance Eq. (\ref{eq:shear_2nd_order}) includes all terms to second order only) leads to a theory which does not have a well posed initial value problem~\cite{Heller:2015dha}. For this reason authors often concern themselves with the M\"uller-Israel-Stewart (MIS) formulation of hydrodynamics~\cite{Muller:1967zza,Israel:1978}. This formulation can be regarded as a UV completion of the standard Landau-Lifschitz formulation of relativistic hydrodynamics in the sense that it describes the dynamics of the system also at very early times~\cite{Heller:2015dha}. In the MIS formulation the shear stress tensor is regarded as a dynamical variable which obeys a relaxation equation,
\begin{equation}
    (\tau_\Pi u^\alpha \partial_\alpha +1)\Pi^{\mu\nu}=-\eta \sigma^{\mu\nu}+\cdots
\end{equation}
where $\tau_\Pi$ is the relaxation time. The solutions for the energy density, or temperature in a boost invariant fluid in MIS theory lead to an infinite series whose radius of convergence is zero. It can then be expected that methods of resurgent analysis may provide further insight into the behavior of the solution. This topic has been explored in a number of publications~\cite{Heller:2015dha,Aniceto:2015mto,spalinskiHydrodynamicAttractorYangMills2018} etc. The MIS equations of motion for our system of interest are given in terms of the dimensionless variables $w=\tau T$ and $f=\tau \dot{w}/w$\footnote{Note that the conventions in which we present the coefficients $C_X$ agrees with~\cite{Romatschke:2017vte}, which deviates from~\cite{Florkowski:2017olj}. However, we have checked that our results are reproduced also when working in the conventions from~\cite{Florkowski:2017olj}.
}
\begin{align}
  0&=  C_{\tau\Pi}f(w) (w f'(w)+4 f(w))+\left(w -\frac{16C_{\tau\Pi}}{3}\right)f(w)\nonumber\\
    &-\frac{4}{9}(C_\eta -4C_{\tau\Pi})-\frac{2}{3}w \, .\label{eq:MIS}
\end{align}
As demonstrated in~\cite{Heller:2015dha} one can look for transseries solutions to Eq. (\ref{eq:MIS}) of the form,
\begin{align}
    f(w)&=\sum_{m=0}^{\infty}c^m\Omega(w)^m\sum_{n=0}^{\infty}a_{n,m}w^{-n}\label{eq:trans} \\
    \Omega(w)&=w^{-\frac{\left(C_\eta-2C_{\lambda 1}\right)}{C_{\tau\Pi}}}e^{-\frac{3}{2}C_{\tau\Pi}w} \nonumber
\end{align}
for which one finds factorially divergent coefficients $a_{n,m}$ (as displayed in Fig.~\ref{fig:Divergent} for the first 250 coefficients of each sector\footnote{It was necessary to use extended precision arithmetic in order to obtain these coefficients, keeping the first 250 decimal places. See provided notebook which computes these coefficients directly.}). 
To compute the Borel resummation one starts with a Borel transformation,  
\begin{equation}
   \hat{f}(\xi)\equiv \mathcal{B}(f(w))=\sum_{n=0}^{\infty}\frac{f_n}{n!} \xi^n \, .
\end{equation}
The Borel transformation can be analytically continued via diagonal Pad\'{e} approximation and the resulting expression transformed back to $w$ via lateral Laplace transformation
\begin{equation}
    \mathcal{L}^{\theta}[\hat{f}](w)=\int_0^{e^{i\theta}\infty}\dd \xi\, e^{-\xi w}\hat{f}(\xi)\,,
\end{equation}
to compute the resummation procedure.

\begin{figure}[t]
\centering
    \includegraphics[width=0.49\textwidth]{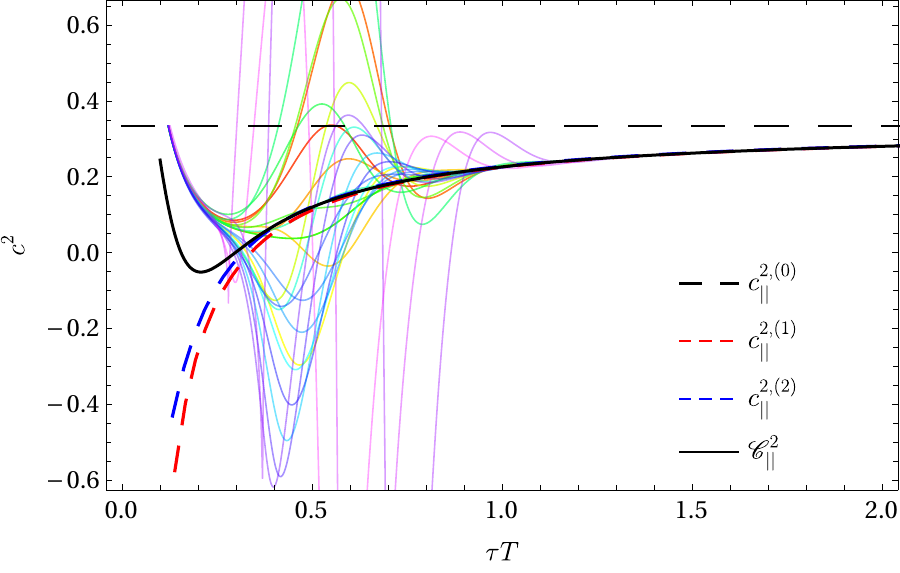}\hfill
    \includegraphics[width=0.49\textwidth]{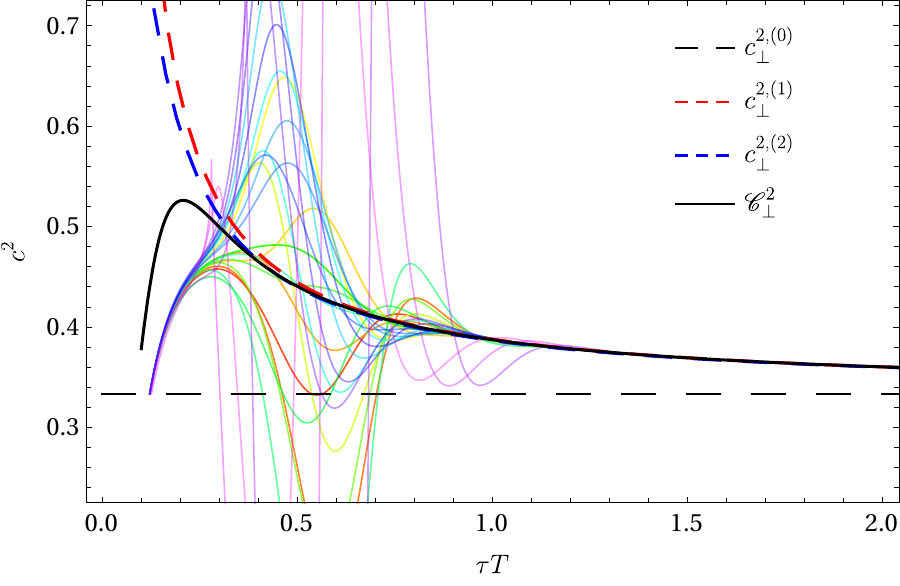}
    \caption{\textit{The speed of sound attractor:} Results for the resummed speed of sound $\mathscr{C}^2$ are displayed. \textit{Left:} $\mathscr{C}_{\para}^2$ \textit{Right:} $\mathscr{C}_{\perp}^2$. Both the results of the hydrodynamic calculation and the out of equilibrium calculation approach the leading order attractor solutions shown as black curves. 
    \label{fig:resum_C}}
\end{figure}

\begin{figure}
    \centering
    \includegraphics[width=0.49\textwidth]{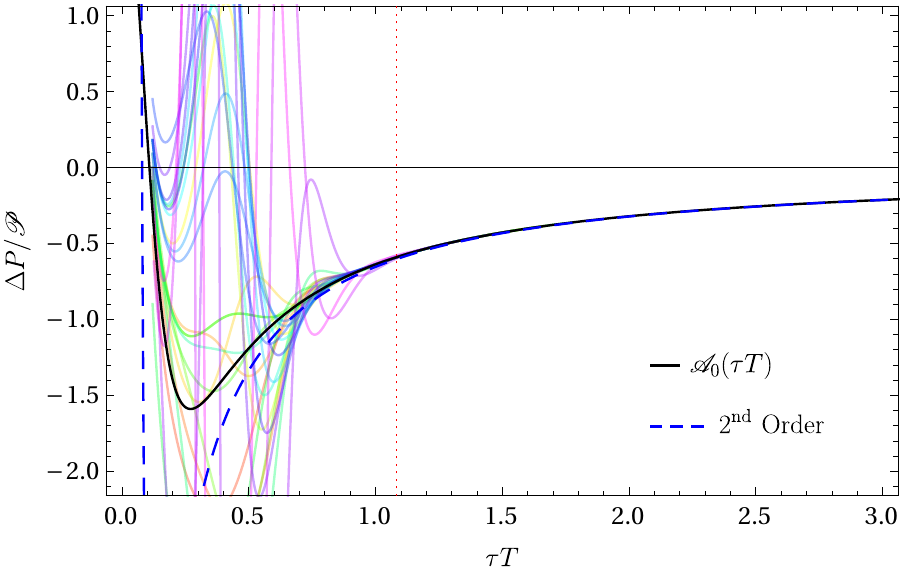}\hfill
    \includegraphics[width=0.49\textwidth]{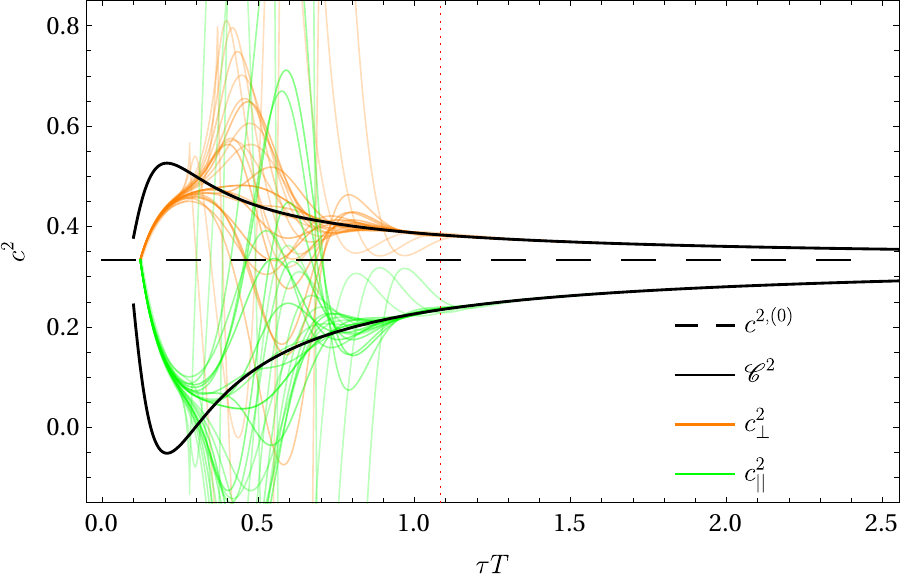}\hfill
    \caption{
    \textit{Attraction:} 
    The attractor behavior of the system is displayed. \Left $\Delta P / \mathscr{P}$ for $\mathscr{P}=\epsilon/3$. \Right the speed of sound $\mathscr{C}^2$ (green longitudinal, orange transverse). The functions in both images are displayed as a function of $\tau T$, where $T=(\epsilon/\sigma_\text{SB})^{1/4}$.
    The vertical dashed red line indicates the approximate onset of an attractor.
    In both images the black lines are the leading order attractor solutions. For $\Delta p / \mathscr{P}$ the blue dashed curve is the attractor as proposed by~\cite{spalinskiHydrodynamicAttractorYangMills2018} given Eq. (\ref{eq:AT}) and the speed of sound attractor solutions, $\mathscr{C}^2$, are given in Eq. (\ref{eq:ATT}) and Eq. (\ref{eq:ATL}).
    Both quantities converge to the attractor at approximately the same time $\tau T\approx 1.1$.
    The horizontal dashed black line shows the conformal value $c^2=1/3$.
    \label{figs:deltaPAndc}}
\end{figure}

In~\cite{spalinskiHydrodynamicAttractorYangMills2018} the author studies the leading order attractor of $\mathcal{N}=4$ SYM theory. Computing 240 coefficients of the solution to Eq.~(\ref{eq:MIS}) the author computed the Borel resummation of the series solution to $f(w)$ and translated the results into the pressure anisotropy
\begin{equation}
     \A(w)=\frac{P_\perp-P_{\para}}{\mathscr{P}}\, , \quad \mathscr{P}=\epsilon/3 \, .
\end{equation}
The resulting resummation is very well approximated by the rational function
\begin{equation}
    \A_0(w)=\frac{2530w-276}{3975 w^2-570w +120}\label{eq:AT} \, .
\end{equation}
This result can be quickly translated into expressions for the speed of sound. To see this, begin by differentiating the pressure anisotropy with respect to the dimensionless time $w$, after some manipulation one finds
\begin{equation}
    \partial_w \A =\frac{\partial_w \epsilon}{\epsilon}\left( 3\Delta c^2 - \A(w)\right)\, ,\quad \Delta c^2=c_{\para}^2-c_{\perp}^2 \, .
\end{equation}
Using the definition of $w$ and the Stefan-Boltzmann relation the ratio $\frac{\partial_w \epsilon}{\epsilon}$ can be re-expressed in terms of $w$ as 
\begin{equation}
    \frac{\partial_w \epsilon}{\epsilon}=\frac{4}{w} \, .
\end{equation}
Combining these two results gives 
\begin{equation}
    \Delta c^2=-\frac{1}{3}\left(\frac{w}{4}\partial_w -1\right)\A(w) \, .\label{eq:Attractor_Pre}
\end{equation}
To extract information about the individual speeds of sound, one can use the trace relation which ensures that $2c_{\perp}^2+c_{\para}^2=1$. A small manipulation reveals that,
\begin{equation}
     \Delta c^2=\frac{-1}{2}+\frac{3}{2}c_{\para}^2=1-3c_\perp^2 \, . 
\end{equation}
With these two results in hand one can isolate $c_\perp^2$ and $c_{\para}^2$ from Eq. (\ref{eq:Attractor_Pre}) which gives
\begin{subequations}
\begin{align}
    \mathscr{C}_{\perp}^2&=\frac{1}{3}+\frac{1}{9}\left(\A_0(w )+\frac{w }{4}  \frac{\partial \A_0(w )}{\partial w }\right) \label{eq:ATT}\, ,\\
    \mathscr{C}_{\para}^2&=\frac{1}{3}-\frac{2}{9} \left(\A_0(w )+\frac{w}{4}  \frac{\partial \A_0(w )}{\partial w }\right) \label{eq:ATL} \, ,
\end{align}
\end{subequations}
where we have replaced $\A$ by its resummed expression $\A_0$ and denoted the resummed speeds of sound by script characters to distinguish them from the other expression used thus far.
Shown in Fig.~\ref{fig:resum_C} is the resummed expression along with the numerical calculation of the out of equilibrium speeds of sound (on the left parallel, on the right transverse). While in Fig.~\ref{figs:deltaPAndc} pressure anisotropy and its attractor are shown on the left, the speeds of sound and their attractor are shown on the right. Finally, it is interesting to note that, to leading order, the attractor behavior of the speed of sounds follows directly from the attractor behavior of the pressure as can be seen by the appearance of $\mathcal{A}_0$ in the expressions for the speed of sound. In addition, not only does the longitudinal direction experience attractor behavior, but also the transverse direction. This is a result of the conformal symmetry which links the evolution of the pressures and the energy density and hence the speeds of sound.

\subsection{Discussion}
\begin{figure}[t]
    \includegraphics[width=0.49\textwidth]{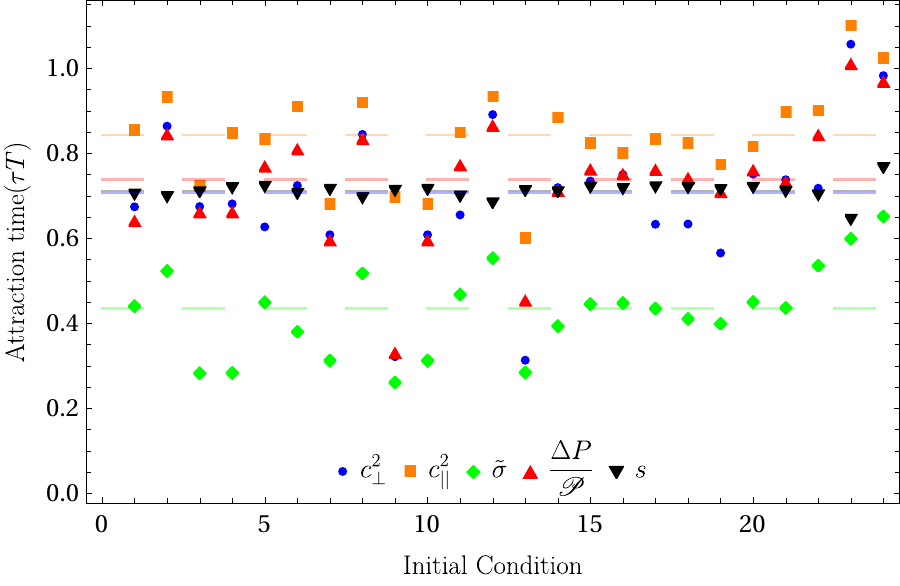} \hfill 
    \includegraphics[width=0.49\textwidth]{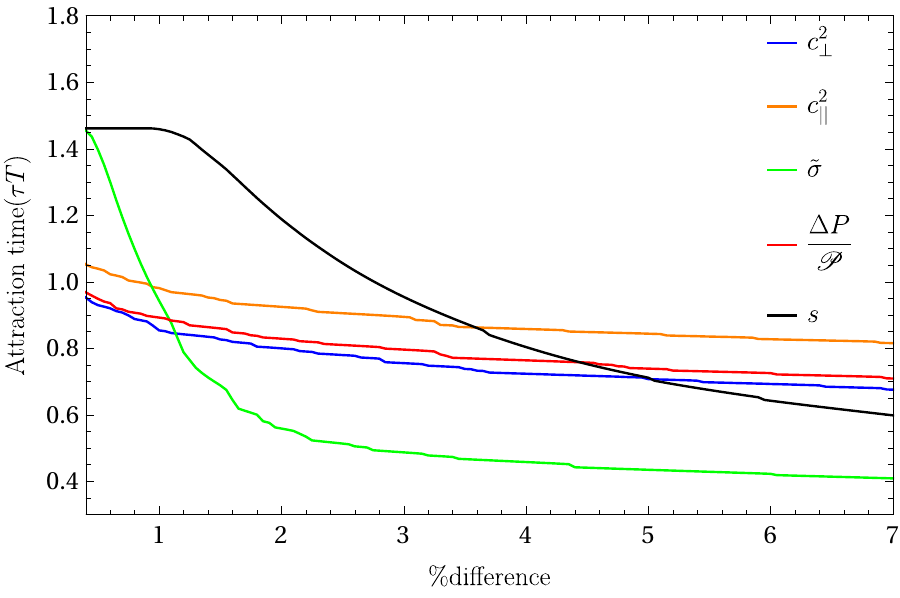}
    \caption{{\it Attractor Behavior:} \Left The times at which the relative difference between the various quantities computed in this work and their attractor is within $5\%$ are displayed. \Right The dependence of the average time on the choice of $\%$ difference. 
    \label{fig:attraction_times}}
\end{figure}

There are four main observations:
\begin{enumerate}
    \item All quantities $f$,\, $c_{\para/\perp},\, \Delta P / \mathscr{P},\, P_{\para}/\epsilon$ and  $\sigma$ quickly approach one \Say{universal}
    curve independent of their initial conditions.
    This occurs approximately at $\tau T\approx 1$ 
    in agreement with other studies~\cite{Heller:2015dha,spalinskiHydrodynamicAttractorYangMills2018,Romatschke:2017ejr}
    \item 
    Interestingly, the only quantity that approaches the attractor not at the same time is the dimensionless entropy density,
    which seems to reach it much earlier\footnote{
  This may not be so surprising given the work of~\cite{Jansen:2016zai,Jansen:2020ign} where the authors find that deformations of the apparent horizon (back-reacted on the geometry) decay with twice the QNM frequency $\propto e^{-2 w_I t}$. 
    }, around $\tau T\approx 0.8$ 
    This has already been mentioned in~\cite{Rougemont:2021qyk}.
    \item The speed of sound calculated with our method also converges to an attractor
    at the same time as the other quantities with the exception of $\sigma$.
    This is not too surprising, since the quantities needed for the calculation of speed of sound are $P_{\perp,\para}$, $\epsilon$ and $\sigma$, which all approach attractors in some way.
    After reaching the attractor the speed of sound only very slowly approaches the
    conformal value of $c^2 = 1/3$.
    There are regions where our notion of speed of sound gives $c_{\para,\perp}^2$ larger than 1 and smaller than 0.
    Both of these regimes indicate instability. Whether this signals a breakdown of our method, or a failure in the ability to interpret our results as a speed of sound requires further analysis. In terms of a mode analysis, $c_{\para,\perp}^2<0$ indicates an instability\footnote{This is often referred to as the gradient instability and is particularly interesting in the non-linear regime of classical EFT models of materials where $c^2<0$ represents the end point of a strain-stress relation. See for example the recent works~\cite{Alberte:2018doe,Pan:2021cux}.}, including an exponentially growing mode. While
    $c_{\para,\perp}^2>1$ clearly violates causality. Interestingly the only curves which violate the causality bound are those which violate the energy conditions (see Fig.~\ref{figs:deltaPandPL}), where we reproduce with our code the corresponding plot from~\cite{Rougemont:2021qyk}.
    Four observations based on our limited data from the six initial conditions which violate the dominant energy condition (DEC), three of which also violate the weak energy condition (WEC):
    \begin{enumerate}
        \item Violation of WEC implies $c_{\para,\perp}^2>1$, the converse is not true.
        \item Violation of the DEC appears to be unrelated to $c_{\para,\perp}^2>1$ (DEC can be violated while $c_{\para,\perp}^2<1$).
        \item Transverse speed of sound: instabilities ($c_{\para,\perp}^2<0$) occur in same cases in which also $c_{\para,\perp}^2>1$.
        \item Longitudinal speed of sound: different from transverse case, it may be that $c_{\para,\perp}^2>1$, but no instability with $c_{\para,\perp}^2<0$ is present.
    \end{enumerate}
    We stress that these are observations based on a small data set, hence, we can not claim but only speculate on these statements being true in general.
    \item
    If any of the pressures should be thought of as a generating functional, then it should satisfy $\epsilon+P=sT$, which the conformal $P$ does the best job at after the least amount of time, then $P_\perp$, then $P_{\para}$ the latest. 
    Considering the pink curve, the attractor in $s$ is reached at $\tau T\approx 0.6$ before the local thermal equilibrium which is reached at $\tau T\approx 0.7$, at which time also the sound attractor is reached, finally the pressure anisotropy attractor is reached the latest at $\tau T\approx 0.9$. This analysis was made more comprehensively by analyzing 25 initial conditions, see Fig.~\ref{fig:attraction_times}.  
\end{enumerate}
In Fig.~\ref{fig:attraction_times} we calculate the time it takes for various quantities to hydrodynamize. This is measured as a percent relative difference,
\begin{equation}\label{eq:relative}
    2\left| \frac{X(\tau T(\tau))-X_\text{expected}(\tau T(\tau))}{X(\tau T(\tau))+X_\text{expected}(\tau T(\tau))}\right| < \delta
\end{equation}
where $100\delta  =\%~\text{difference}$. A similar study was conducted in~\cite{Rougemont:2021gjm} of the hydrodynamization times in the same system. It is important to note a few relevant differences. First we do not the same normalization used in~\cite{Rougemont:2021gjm}, that is an equation of the form
\begin{equation}
    \left| \frac{X(\tau T(\tau))-X_\text{expected}(\tau T(\tau))}{X_\text{expected}(\tau T(\tau))}\right| < \delta
\end{equation}
as we feel this should only be used when there is a ``correct'' answer. 
In our case, however, the quantities we compare to are approxmiations, such as the second order hydrodynamic values of pressures or energy density. 
Furthermore, the Borel resummed expressions we have discussed are only the leading order solutions. In addition, we find it misleading to 
normalize every quantity to an 
effective temperature, 
especially when the effective temperature is a truncated solution to the hydrodynamic equations. 
For hydrodynamic quantities we normalize to the temperature expanded to the same order; e.g. $\sigma_{\text{2nd}}/T_{\text{2nd}}$. 
Any time we reference quantities computed using the gravitational evolution we use the full effective temperature as given by the Stefan-Boltzmann equation, for example we use $\sigma = s(\tau)/T^3$ where $T$
is given in terms of the gravitational evolution as $T=(\epsilon/\sigma_\text{SB})^{1/4}$. Each relative difference can be computed using these definitions to find a unique $\tau_*$ for which the expression in Eq.~(\ref{eq:relative}) is satisfied for all future times. Given such a $\tau_*$ the values reported in the figure are constructed with the full effective temperature, i.e.\ $\tau_*T(\tau_*)$

In Fig.~\ref{fig:attraction_times} we have compared each expression to its best case scenario, i.e.\
\begin{itemize}
    \item The speeds of sound to their Borel resummed expressions. 
    \item The pressure anisotropy to its Borel resummed expression.
    \item The entropy density as calculated from the apparent horizon to the second order hydrodynamic approximation of the field theory entropy density.
    \item The entropy density as calculated from the apparent horizon to the entropy density extracted from the Euler relation. 
\end{itemize}

Figure~\ref{fig:attraction_times} shows on the left plot the times at which the relative difference between the various quantities computed in this work and their attractor is within $5\%$. 
The dashed lines indicate the mean time of attraction. The entropy first reaches its attractor behavior followed by a local notion of thermal equilibrium. Shortly after the perpendicular speed of sound reaches its attractor, then the pressure anisotropy reaches its attracting behavior followed by the longitudinal speed of sound. Notice that there are initial conditions for which this ordering, based on the average time of attraction, is not obeyed.
We also point out that these attractor times can not be trusted below 2 percent threshold, as visualized in the right side plot in Fig.~\ref{fig:attraction_times}. That right plot shows the dependence of the average time on the choice of $\%$ difference, i.e.\ the threshold we choose. The average attraction time shows a mild dependence on the choice of acceptance percentage. 
It is interesting to note that the entropy approaches the hydrodynamic expectation but never truly makes it to the thermal expectation, it always stays a distance away from its local thermal equilibrium value. I.e.\ there is never a local thermal equilibrium (with the entropy given by the horizon area) despite hydrodynamics being a good description.  

\section{Conclusions}
\label{sec:conclusions}
In this work, we have proposed a method to compute the speed of sound out of equilibrium, Eq. \eqref{eq:Speed_of_sound_NONEQ}, in a conformal fluid, and compared it to three other methods, enumerated in section~\ref{sec:out_of_eq}, see Fig.~\ref{fig:Methods_Comparison}. While methods 1, 2, and 3 lead to superluminal speeds, method 4 is well-behaved. Merely in the pathological cases for fluids disobeying the weak energy condition, see for example the blue curve in Fig.~\ref{figs:deltaPandPL}, method 4 would yield superluminal speeds of sound. This method computes the speeds, Eq.~\eqref{eq:outOfEqCs} by varying the pressure and energy density at a fixed proper time. Results we obtain include: 
\begin{itemize}
    \item The hydrodynamic expectation for out-of-equilibrium speed of sound was computed to second order in viscous derivative corrections. An analytic form for the speed of sound along the Bjorken expansion is given in Eq.~\eqref{eq:cperp_hydro}, and transverse to it in Eq.~\eqref{eq:cpara_hydro}. 
    \item The out-of-equilibrium speed of sound was computed within a holographic model, namely Bjorken-expanding $\mathcal{N}=4$ SYM plasma. Note, that the holographic time-evolution is well approximated by the second order hydrodynamic prediction, Fig.~\ref{fig:hydro_comp}. The deviation from the conformal value 1/3 is dominated by the shear stress, $\pi_L$, see Fig.~\ref{fig:shear_contributions_isolated}. 
    \item The out-of-equilibrium speed of sound attractors for sound propagation longitudinal and transverse to the beamline were computed analytically, see Eq.~\eqref{eq:ATL} and \eqref{eq:ATT}, respectively. 
    \item Entropy density of the field theory computed from the apparent horizon area, Eq.~\eqref{eq:entropy}, normalized to the temperature defined through the fourth root of the energy density, as displayed in Fig.~\ref{fig:Scaled_Entropy}. This dimensionless entropy density reaches an attractor, see Fig.~\ref{fig:Scaled_Entropy} (in contrast to the entropy density normalized to $T_{ideal}$~\cite{Rougemont:2021gjm,Rougemont:2021qyk}). In fact, this entropy density is the first of all quantities to reach an attractor as seen by the average lines in Fig.~\ref{fig:attraction_times}. 
    \item Local thermodynamic consistency, $\epsilon(x)+P(x)= s(x) T(x)$, was used to define local thermal equilibrium. The time scale at which this condition is satisfied was compared at a 5\%-threshold to the times at which different initial conditions reach the hydrodynamic entropy attractor, the anisotropy attractor, and the two sound attractors, see Fig.~\ref{fig:attraction_times}. On average, the apparent horizon entropy density reaches the attractor first ($\tau T\approx 0.43$), followed by the transverse speed of sound simultaneously with the establishment of local thermal equilibrium ($\tau T\approx 0.71$), then the pressure anisotropy reaches its attractor ($\tau T\approx 0.74$) and finally the longitudinal speed of sound ($\tau T\approx 0.85$). 
    However, as seen from the right plot in Fig.~\ref{fig:attraction_times}, it is not clear that local thermal equilibrium is ever reached below the 2\%-threshold.
    \item Out of our 25 initial conditions, six are violating the DEC, three of which are also violating the WEC. Within this limited set we observe that a violation of the WEC implies causality violation by the speeds of sound $c_{\para,\perp}^2>1$. 
\end{itemize}

Our results confirm the statement~\cite{Kurkela:2019set} that in strongly coupled systems hydrodynamic attractors are reached after the system follows hydrodynamic time-evolution equations, i.e.\ after hydrodynamization. Furthermore, we confirm that local thermal equilibrium is neither required for, nor implied by reaching a hydrodynamic attractor, see Fig.~\ref{fig:attraction_times}. 

Because the entropy reaches the attractor at much earlier times than the other quantities like pressure anisotropy or $f$, one wonders about the reason. 
We have defined the entropy through the area of the apparent horizon~\cite{Bhattacharyya:2008xc,Booth:2010kr,Booth:2011qy,Engelhardt:2017aux}.
One potential resolution is that the information of the change of the apparent horizon area has to propagate to the boundary. This could lead to a delay before the field theory learns about the corresponding change in entropy. 
For this reason, in order to get an estimate for the time it takes to propagate information, we calculated lightlike geodesics from the apparent horizon to the boundary. It turns out that the delay is of the right order of magnitude $\mathcal{O}(1)$, 
but taking into account the appropriate time delay for each apparent horizon area destroys the attractor behavior in the resulting putative entropy measure, leaving the appropriate field theory entropy measure as an open question. 

As a next step and rigorous check of our proposal, we intend to compute the speed of sound directly from the scalar fluctuations (spin-0 under rotations), including the sound channel fluctuations around the time-dependent background metric~\eqref{eq:line_element}, which we have analyzed here. By comparison to the result of that \emph{in situ} computation of the (true) speed of sound to the out-of-equilibrium speed of sound we propose here based on energy density, diagonal components of the energy-momentum tensor and entropy, this will reveal the validity of this method. However, this is a challenging distinct computation, which is why we defer it to later work. 

It should be noted that the simple idea of a sound wave propagating through a time-dependent medium may be made rigorous in the context of the Schwinger-Keldysh formulation of hydrodynamics and its stochastic corrections on the level of a generating functional~\cite{Haehl:2015pja,Haehl:2015foa,Crossley:2015evo,Glorioso:2018wxw}. Within this formulation, interactions between hydrodynamic fluctuations are taken into account, and it may be possible to derive corrected eigenmode equations for fluctuations around Bjorken flow within this framework. These may assume the form of wave equations and the speed of the wave may be determined by variations of the generating functional with respect to hydrodynamic fields, similar to $\partial P/\partial \epsilon$, where $P$ should be viewed as the relevant generating functional~\cite{Kovtun:2016lfw,Hernandez:2017mch}. 
One strong indication that this realization within then Schwinger-Keldysh formulation (or in a more general far-from-equilibrium fluid description~\cite{Romatschke:2017vte}) must be possible, is the fact that the dual gravitational action does exactly that: it serves to derive fluctuation equations around any given metric background, including the Bjorken-expanding five-dimensional metric we worked with in this paper. This time-dependent metric background is dual to the Bjorken-expanding plasma and a subset of the gravitational fluctuations are longitudinal, including those eigenmodes which turn into the sound modes in the hydrodynamic regime. When the system is outside the hydrodynamic regime, far from equilibrium, those modes should still be eigenmodes and propagate with a speed which we may consider the out-of-equilibrium speed of sound. For a very recent holographic study of correlation  functions in this context, see~\cite{Banerjee:2022aub}.

Conformal symmetry relates all quantities to the time-evolution of the energy density in a Bjorken expanding plasma like the one discussed in this work. Remarkably, much less symmetric examples of non-conformal systems (introducing massive particles) still show early-time attractor behavior in the longitudinal pressure. This behavior is not matched by hydrodynamics around an isotropic equilibrium state, however, it can be matched by an \emph{anisotropic hydrodynamics} description~\cite{Jaiswal:2021uvv}, underlining the importance of developing anisotropic descriptions of heavy ion collisions~\cite{Romatschke:2003ms,Ryblewski:2010tn,Strickland:2014pga,Florkowski:2012lba,Ammon:2017ded,Ammon:2020rvg,Garbiso:2020puw,Cartwright:2021maz,Cartwright:2021qpp}.

Finally, the obvious question is, what relevance our work has for the further exploration of the QCD critical point and the QCD equation of state\cite{Parotto:2018pwx,parottoQCDEquationState2020,Martinez:2019bsn}. What happens to the speed of sound when both conditions are met, out-of-equilibrium and proximity to the QCD critical point? 
These questions are also under investigation in the beam energy scan~\cite{Odyniec:2019kfh,BES_WPII,Bzdak:2010fd} and the beam energy scan theory initiative~\cite{An:2021wof}. 
A holographic investigation of this may be possible by combining our approach in this paper with techniques from Hydro+~\cite{Stephanov:2017ghc} (see also~\cite{Abbasi:2021rlp}) on the field theory side, a critical point in the holographic model~\cite{DeWolfe:2010he,Critelli:2017euk}, and most importantly including conserved charges to our analysis (attractors in a hydrodynamic model including baryon charge for instance have been studied in~\cite{Dore:2020jye}). The possible measurement of the speed of sound provided through the measurement of baryon cumulants as proposed in~\cite{Sorensen:2021zme} could offer access to holographic out-of-equilibrium speeds of sound for comparison to the expectations of the heavy ion community.


\acknowledgments 
We thank Ulrich Heinz, Juan Hernandez, Jakub Jankowski, Thomas Sch\"afer, Bj\"orn Schenke and Michal Spalinski for discussions, as well as Willians Barreto, R\^omulo Rougemont and Jorge Noronha for discussing data shared from their publication~\cite{Rougemont:2021qyk}. For comments on this manuscript we are grateful to S\"oren Schlichting, Matteo Baggioli and Michal Spalinski. This work was supported, in part, by the U.S.~Department of Energy grant DE-SC-0012447. CC also acknowledges support by the Netherlands Organisation for Scientific Research (NWO) under the VICI grant VI.C.202.104.

\bibliography{bib.bib}


\onecolumngrid
\appendix
\section{Further details on the numerics} \label{appendix:Numerics}
Using these new \Say{dotted} derivatives, the Einstein Field Equations take the form
\begin{subequations}
    \begin{align}
        S''(v,r)&=-\frac{1}{2} B'(v,r)^2 S(v,r) \label{einS} \, ,\\
        \dot{S}'(v,r)&=-\frac{2 S'(v,r)  \dot{S}(v,r)}{S(v,r)}+2 S(v,r)\label{einSd} \, ,\\
        \dot{B}'(v,r)&= -\frac{3 \dot{B}(v,r) S'(v,r)}{2 S(v,r)}-\frac{3 B'(v,r)
        \dot{S}(v,r)}{2 S(v,r)}+\label{einBdot}\, ,\\
        A''(v,r)&= -3 B'(v,r) \dot{B}(v,r)+\frac{12 S'(v,r) \dot{S}(v,r)}{S(v,r)^2}-4\label{einA}\, ,\\
        \ddot{S}(v,r)&= \frac{1}{2} A'(v,r) \dot{S}(v,r)-\frac{1}{2} \dot{B}(v,r)^2 
        S(v,r).\label{einsdotdot}
    \end{align}
    \label{eq:EFEs}
\end{subequations}
The revelation is that this set of differential equations is nested\footnote{We also wrote a Mathematica package to visualize the structure of differential equations and see the nested structure, see \url{https://github.com/BoGGoG/DEQSystemStructureVisualization}}, i.e.\ 
for a given $v_0$, one can start from the first DEQ~(\ref{einS}), using some $B(v_,r)$ as initial condition,
and solve one's way from equation to equation in order to get $S(v_0, r)$, $\dot{S}(v_0, r)$, $\dot{B}(v_0, r)$, and $A(v_0, r)$.
Note that we solve for $\dot{S}$ and $S$ independently as well as $\dot{B}$ and $B$.
The last equation, equation~(\ref{einsdotdot}) is not needed and can be used as a constraint equation to check the numerics.

\subsection{Residual gauge freedom}
It turns out that the Einstein Field equations in~(\ref{eq:EFEs}) possess a residual gauge freedom related to bulk diffeomorphisms, namely
\begin{equation}
    r \to r + \lambda(v)\,.
    \label{eq:radial-shift}
\end{equation}
One could just randomly set $\lambda$ to some constant, but there is actually a better choice by~\cite{Chesler:2013lia} and really well explained in~\cite{vanderSchee:2014qwa}: 

We go from the radial coordinate $r$ to its inverse $z=1/r$.
We have not yet specified the grid for the bulk integration in order to solve equations~(\ref{eq:EFEs}) numerically.
It is obvious that one end of the interval of integration should be the boundary, $r_\text{bdy}\to \infty$ or in practice $z_\text{bdy}=1/r_\text{bdy}=0$.
The other end of the interval is not so trivial.
Ideally we should integrate exactly to the event horizon, but the event horizon is a teleological
object, i.e.\ in order to know where it is we would need to know the future evolution~\cite{Poisson:2009pwt}.
Also, $S$ usually vanishes at some point, which means there is a caustic.
It turns out that usually this caustic is hidden behind the event horizon, but it can state
a limitation on the initial conditions.
For our metric and coordinates the apparent horizon is determined by
\begin{equation}
    \dot{S}(v,z_\text{h})=0.
    \label{eq:apparent-horizon}
\end{equation}

While the event horizon cannot be determined prior to knowing the whole evolution,
the \textit{apparent horizon} can be determined for every time slice, lies inside the event horizon
and converges to the event horizon for late times.
Thus, the procedure will be to use the radial shift invariance~(\ref{eq:radial-shift}) to set the apparent horizon to $r=1$ for all times.
We calculate $\lambda$ for the initial time and $B_s$ profile and from the subtraction scheme of $A$ (explained in 
section~\ref{subsec:Subtractions})
we have an equation for $\mathrm{d}\lambda/\mathrm{d} v$, so we can take $\lambda$ as another variable in
$\Phi$ that is propagated from time slice to time slice. 

This way we can always integrate on the interval $z \in [0,1]$.
The numerical integration on this interval is performed using \textit{Pseudospectral methods}\footnote{We also wrote a Mathematica package for this, see \url{https://github.com/BoGGoG/MathematicaChebyshevSolver}, even though many changes have not yet been pulled into this repo.} following~\cite{Boyd00}.
For the grid size we use 36 points.


\subsection{Obtaining regular functions}
\label{subsec:Subtractions}
One practical problem of the Einstein Field Equations in the form above is that the functions
$A$, $B$, $S$ generally diverge at the boundary.
This problem can be circumvented by pulling out the divergences from those functions and only solving for the regular part.
This can be achieved with the following choice of \Say{subtraction scheme} (now written in terms of $z=1/r$):
\begin{subequations}
    \begin{align}
      A(v,z)&=z^2 A_s(v,z)+\lambda (v)^2+\frac{2 \lambda (v)}{z}+\frac{1}{z^2}\,,\\
       B(v,z)&=z^4 B_s(v,z)-\frac{2 z^3 \left(3 v^2 \lambda (v)^2+3 v \lambda (v)+1\right)}{9 v^3}+\frac{z^2 (2 v \lambda (v)+1)}{3 v^2}-\frac{2 z}{3 v}-\frac{2 \log (v)}{3}\,, \\
       S(v,z)&=z^3 S_s(v,z)+\frac{3 v \lambda (v)+1}{3 v^{2/3}}+\frac{z^2 (9 v \lambda (v)+5)}{81 v^{8/3}}-\frac{z}{9 v^{5/3}}+\frac{v^{1/3}}{z}\,.
    \end{align}
    \label{eq:SubtractionScheme}
\end{subequations}
We call the functions $A_s$, $B_s$ and $S_s$ the \textit{subtracted functions} and they are scaled by factors of $z$ because this will be the first order where they are non-zero.
After plugging equations~(\ref{eq:SubtractionScheme}) into the EFEs~(\ref{eq:EFEs}), we have a nested set of differential equations for the functions $A_s$, $B_s$ and $S_s$ that we can solve just the way that has been indicated below equations~(\ref{eq:EFEs}).

\subsection{Energy-momentum tensor}
From the holographic equation for the energy-momentum tensor~(\ref{eq:Tmunu0}) and using our metric~(\ref{eq:line_element})
we can get the energy-momentum tensor.
It is diagonal and traceless with components
\begin{align}
 \bar{\kappa}\vev{T^0_0}&=\frac{3}{4} a_4(\tau)\label{eq:energy} \\
 \bar{\kappa} \vev{T^{1}_1}=\bar{\kappa}  \vev{T^{2}_2}&=b_4(\tau)-\frac{3 \tau^4 a_4(\tau)+4 \tau \lambda (\tau) (\tau \lambda (\tau) (2 \tau \lambda (\tau)+3)+2)+2}{12 \tau^4} \label{eq:PT} \\
 \bar{\kappa} \tau^2\vev{T^{\xi}_\xi}&= -\frac{1}{4} \tau^2 \left(a_4(\tau)+8 b_4(\tau)\right)+\frac{1}{3 \tau^2}+\frac{4}{3} \tau \lambda (\tau)^3+2 \lambda (\tau)^2+\frac{4 \lambda (\tau)}{3 \tau} \label{eq:PL}
\end{align}
with the transverse coordinates $x_1$, $x_2$ and the (longitudinal) rapidity $\xi=\frac{1}{2} \ln [(t+x_3)/(t-x_3)]$.
$\bar{\kappa} = 4 \pi G_N$ is a normalization constant.
The terms $a_4$ and $b_4$ are the 4th order coefficients of the expansion of $A$ and $B$ at the boundary in terms of $r$.
From the energy-momentum tensor we can read off the energy density $\epsilon = \vev{T_{\tau}^\tau}$
and the transverse and longitudinal pressures $P_\perp =  \vev{T_{x_1}^{x_1}} = \vev{T_{x_2}^{x_2}}$, $P_{\para} = \vev{T^\xi_\xi}$.

\subsection{Initial Conditions}
For completeness we display the initial conditions chosen for this work. This is in part a reproduction of the table in~\cite{Rougemont:2021qyk}.

\begin{table}[h]
 \caption{\textit{Initial Conditions:} The different values of the parameterization of the initial data given in Eq. (\ref{eq:IC_Parameterization}) are displayed. For each parameterization we begin the evolution at $\tau=0.2$ and with initial asymptotic coefficient $a_4=-40/3$ except for initial conditions 24 and 25 for which $a_4=-15.5$ and $a_4=-14.2$ respectively. Note, we do not alter $\alpha$ from the value $\alpha=1$. Doing so leads to initial conditions which do not initially begin as deviations on top of a vacuum AdS solution.  \label{tab:IC}}
 \begin{ruledtabular}
 \begin{tabular}{llllllllllllll}
 IC \# & $\Omega_1$ & $\gamma_1$ & $\Omega_2$ & $\gamma_2$ & $\Omega_3$ & $\gamma_3$ & $\beta_0$ & $\beta_1$ & $\beta_2$ & $\beta_3$ & $\beta_4$ & $\beta_5$ & $\alpha$  \\
1& 0 & 0 & 0 & 0 & 0 & 0 & 0.5 & -0.5 & 0.4 & 0.2 & -0.3 & 0.1 & 1\\
2& 0&0 &0 &0 &0 &0& 0.2 & 0.1 & -0.1 & 0.1 & 0.2 & 0.5 & 1 \\
3&0&0&0&0&0&0& 0.1 & -0.5 & 0.5 &0&0&0& 1\\
4&0&0&0&0&0&0& 0.1 & 0.2 & -0.5 &0&0&0& 1\\
5&0&0&0&0&0&0& -0.1 & -0.4 &0&0&0&0& 1\\
6&0&0&0&0&0&0& -0.2 & -0.5 & 0.3 & 0.1 & -0.2 & 0.4 & 1\\
7&0&0&0&0&0&0& 0.1 & -0.4 & 0.3 &0& -0.1 &0& 1\\
8&0&0&0&0&0&0&0& 0.2 &0& 0.4 &0& 0.1 & 1\\
9&0&0&0&0&0&0& 0.1 & -0.2 & 0.3 &0& -0.4 & 0.2 & 1 \\
10&0&0&0&0&0&0& 0.1 & -0.4 & 0.3 &0& -0.1 &0& 1 \\
11& 1& 1&0&0&0&0&0&0&0&0&0&0& 1\\
12&0&0& 1& 1&0&0&0&0&0&0&0&0& 1\\
13&0&0&0&0&0&0& 0.1 & -0.4 & 0.4 &0& -0.1 &0& 1\\
14&0&0&0&0&0&0& -0.2 & -0.5 & 0.3 & 0.1 & -0.2 & 0.3 & 1 \\
15&0&0&0&0&0&0& -0.2 & -0.3 &0&0&0&0& 1\\
16&0&0&0&0&0&0& -0.2 & -0.5 &0&0&0&0& 1\\
17&0&0&0&0&0&0& -0.1 & -0.3 &0&0&0&0& 1\\
18&0&0&0&0&0&0& -0.1 & -0.2 &0&0&0&0& 1\\
19&0&0&0&0&0&0& -0.5 & 0.2 &0&0&0&0& 1\\
20&0&0&0&0&0&0& -0.2 & -0.4 &0&0&0&0& 1\\
21&0&0&0&0&0&0& -0.2 & -0.6 &0&0&0&0& 1\\
22&0&0&0&0&0&0& -0.3 & -0.5 &0&0&0&0& 1\\
23&0&0&0&0& 1& 8. &0&0&0&0&0&0& 1\\
24& 1& 8 &0&0&0&0& -0.2 & -0.5 &0&0&0&0& 1\\
25&0.5 & 8 &0&0&0&0& -0.2 & -0.5 &0&0&0&0& 1\\
\end{tabular}
\end{ruledtabular}
\end{table}

\section{Horizon fixing schemes}
\label{append:Horizon_Fixing}
In this appendix we display three different methods of running our numerical code, displaying that our evolution procedure provides identical results to~\cite{Rougemont:2021qyk}. We begin with the following proposition,
\begin{itemize}
    \item[\textit{Proposition:}] The triple $(B(v_0,z),\epsilon(v_0), \lambda(v_0))$ is a representative of a class of initial data. This class of initial data is gauge equivalent to all other choices of initial data related by a radial diffeomorphism 
\begin{equation}
    z'=\frac{z}{1+\lambda' z}
\end{equation}
Any choice related by the above diffeomorphism represents equivalent initial data. 
\end{itemize}
Consider the line element for this setup given in Eq. (\ref{eq:line_element}). 
Solving the Einstein equations order by order the near the AdS boundary can be done by making the ansatz
\begin{equation}
    g_{\mu\nu}=\sum_{n=0}^\infty g^{(n)}_{\mu\nu}(x^i)r^{2-n}
\end{equation}
and results in the following expansion for the metric components $B$ 
\begin{equation}\label{eq:expand_b}
    B=\frac{b_4(v)}{r^4}-\frac{-2 v \lambda (v)-1}{3 r^2 v^2}-\frac{12 v^2 \lambda (v)^2+12 v \lambda (v)+4}{18 r^3 v^3}-\frac{2}{3 r v}+\log \left(\frac{1}{v^{2/3}}\right)+\cdots \, ,
\end{equation}
and $A$ 
\begin{equation}
   A= \frac{a_4(v)}{r^2}+r^2+2 r \lambda (v)+\lambda (v)^2-2\lambda'(v) +\cdots \, ,
\end{equation}
in a general frame. 
In this frame one can construct the energy-momentum tensor following~\cite{Skenderis:2008dg,Taylor:2000xw,Fuini:2015hba} to obtain the result displayed in Eq.~(\ref{eq:energy})-(\ref{eq:PL}) which contains explicit factors of $\lambda$, the residual diffeomorhpism left over from the choice of metric ansatz. Constructing a physical observable, the pressure anisotropy per energy density say, one again sees the presence of $\lambda$
\begin{equation}\label{eq:press_anis}
\frac{\Delta P}{\epsilon}=\frac{P_\perp-P_{\para}}{\epsilon}=-\frac{-12 \tau^4 b_4(\tau)+8 \tau^3 \lambda (\tau)^3+12 \tau^2 \lambda (\tau)^2+8 \tau \lambda (\tau)+2}{3 \tau^4 a_4(\tau)}
\end{equation}
However the conclusion that the expression above containing $\lambda$ implies the system is gauge dependent is not correct. To see this lets return to the expansion in Eq.~(\ref{eq:expand_b}). The energy-momentum tensor was computed in this frame, so $b_4$ in Eq.~(\ref{eq:press_anis}) is the same as in Eq.~(\ref{eq:expand_b}). While it appears that $b_4$ is gauge independent this is not true. This can be seen easily by taking $\lambda=0$ in Eq.~(\ref{eq:expand_b}) and then performing a transformation back to the $\lambda$ frame by $r\rightarrow r+\lambda$
\begin{equation}
   B= -\frac{1}{3} (2 \log (\tau))-\frac{2}{3 r \tau}+\frac{2 \tau \lambda (\tau)+1}{3 r^2 \tau^2}-\frac{2 \left(3 \tau^2 \lambda (\tau)^2+3 \tau \lambda (\tau)+1\right)}{9 r^3 \tau^3}+\frac{b_4(\tau)+\frac{\lambda (\tau)^2}{\tau^2}+\frac{2 \lambda (\tau)}{3 \tau^3}+\frac{2 \lambda (\tau)^3}{3 \tau}}{r^4}+\cdots
\end{equation}
We can now see that the 4th order coefficient clearly displays gauge dependence. If we do another transformation $r\rightarrow r+\lambda_2$ we find,
\begin{equation}
   \frac{b_4(\tau)+\frac{\lambda (\tau)^2}{\tau^2}+\frac{2 \lambda (\tau)}{3 \tau^3}+\frac{2 \lambda (\tau)^3}{3 \tau}}{r^4}\rightarrow \frac{b_4(\tau)+\frac{(\lambda (\tau)+\lambda_2 (\tau)) (\tau (\lambda (\tau)+\lambda_2 (\tau)) (2 \tau (\lambda (\tau)+\lambda_2 (\tau))+3)+2)}{3 \tau^3}}{r^4}
\end{equation}
Clearly the gauge invariant contribution to the fourth order coefficient, $b_4$, is the quantity
\begin{equation}
    B_4(\tau)=b_4(\tau)-\frac{\lambda (\tau)^2}{\tau^2}-\frac{2 \lambda (\tau)}{3 \tau^3}-\frac{2 \lambda (\tau)^3}{3 \tau}
\end{equation}
In the case of $A$ we find that $a_4\rightarrow a_4$ and hence is un-effected by gauge transformations. Inserting the gauge invariant fourth order expansion coefficient $B_4$ into the pressure anisotropy equation reveals,
\begin{figure}[t]
    \centering
    \includegraphics[width=0.49\textwidth]{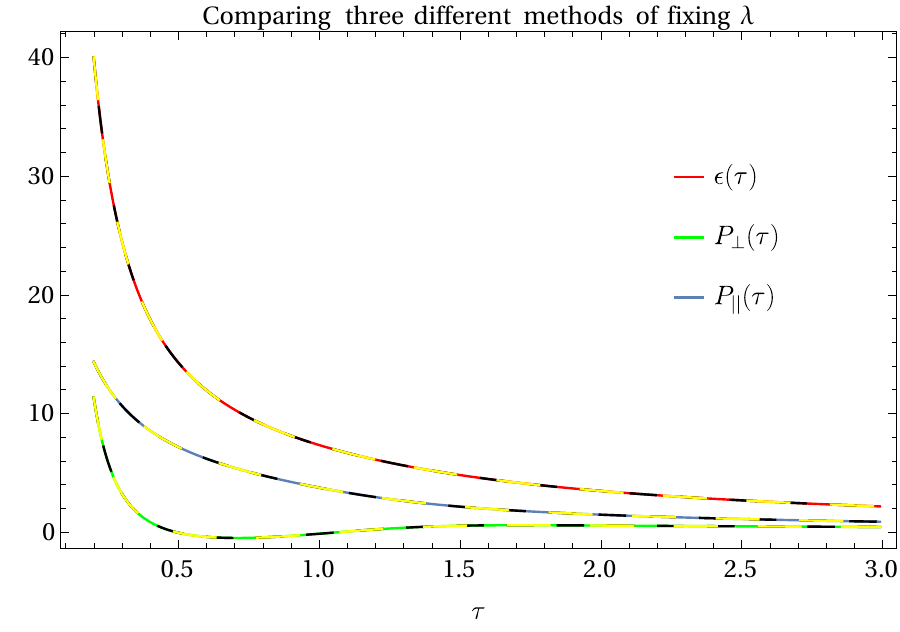}\hfill     \includegraphics[width=0.49\textwidth]{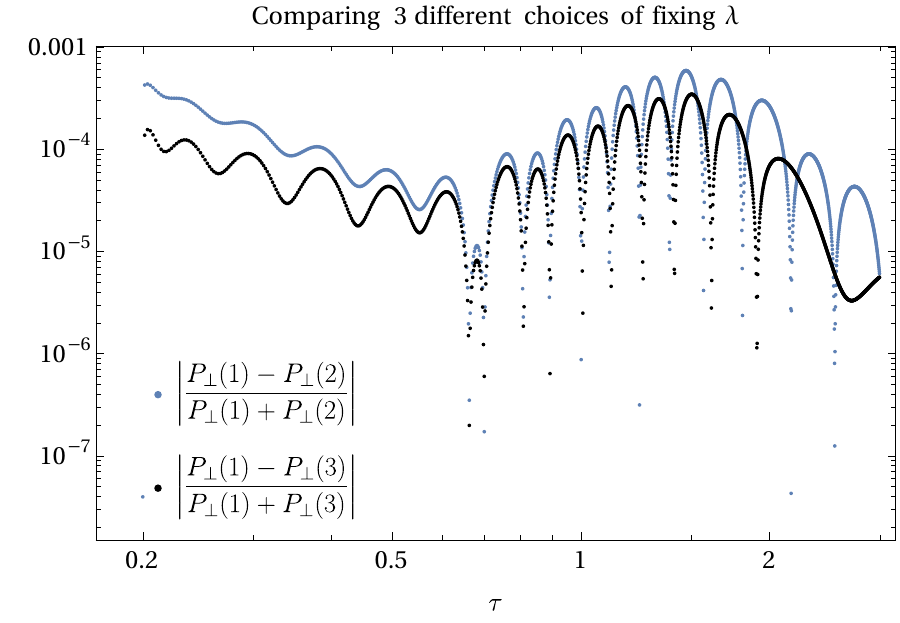}
    \caption{\Left Gauge equivalent evolution. \Right Difference in $P_\perp$. The difference is at the same order as the accuracy to which the apparent horizon stayed at the correct location. The numbers correspond to the method as enumerated in the main text.
    \label{fig:Geq}}
\end{figure}
\begin{equation}
  \frac{\Delta P}{\epsilon}=  \frac{2 \left(6 B_4(\tau) \tau^4-1\right)}{3 \tau^4 a_4(\tau)}
\end{equation}
 which under a radial gauge transformation is invariant. The energy-momentum tensor written in terms of the invariant information takes the form,
\begin{align}
     \bar{\kappa}\vev{T^0_0}&=\frac{3}{4} a_4(\tau) \\
 \bar{\kappa} \vev{T^{1}_1}=  \bar{\kappa}\vev{T^{2}_2}&=-\frac{a_4(\tau)}{4}+B_4(\tau)-\frac{1}{6 \tau^4}\\
 \bar{\kappa} \tau^2\vev{T^{\xi}_\xi}&= -\frac{1}{4} \tau^2 a_4(\tau)-2 B_4(\tau) \tau^2+\frac{1}{3 \tau^2}
\end{align}
Hence we see an under appreciated aspect of this characteristic evolution procedure
 \begin{itemize}
     \item[] \textit{The na\"ive $b_4(t)$ is a gauge dependent quantity and the presence of the $\lambda$ in the energy-momentum tensor is there, precisely, to cancel this gauge dependence. }
 \end{itemize}
It is often the case that the energy-momentum tensors obtained in terms of asymptotic coefficients displayed in other manuscripts which use the characteristic formulation are displayed in the $\lambda=0$. This is no coincidence as the for of the energy-momentum tensor in the $\lambda=0$ is identical to the form in the gauge invariant frame. To further prove this point we have constructed three evolutions, each conducted with $n_z=35$ grid points and $\Delta v=8.33\times 10^{-5}$
\begin{enumerate}
    \item Colored lines represent $\lambda(v_0)\neq 0$ chosen such that $z_h=1$ and is fixed throughout the evolution such that (within approximately one part in $10^{-7}$ which is one notion of the accuracy of the solution), $z_h=1$ for the entire evolution.
    \item Black dashed lines represent $\lambda(v)=0 \,\forall \, v$ . Hence the location of the apparent horizon fluctuates throughout the evolution. 
    \item Yellow dashed lines represent $\lambda(v_0)\neq 0$ for which the apparent horizon is at $z_h$. $\lambda$ is then fixed throughout the evolution such that (within approximately one part in $10^{-7}$ which is one notion of the accuracy of the solution), $dz_h/dv=0$ for the entire evolution. 
\end{enumerate}
One can note that all 3 curves are visually identical and differ from one another on the order of $10^{-4}$. This is an example of the proposition stated above. All three of these evolutions belong to the same class of initial data, they are all gauge equivalent to one another.

\section{Energy Conditions}
\begin{figure}[t]
    \centering
    \includegraphics[width=0.49\textwidth]{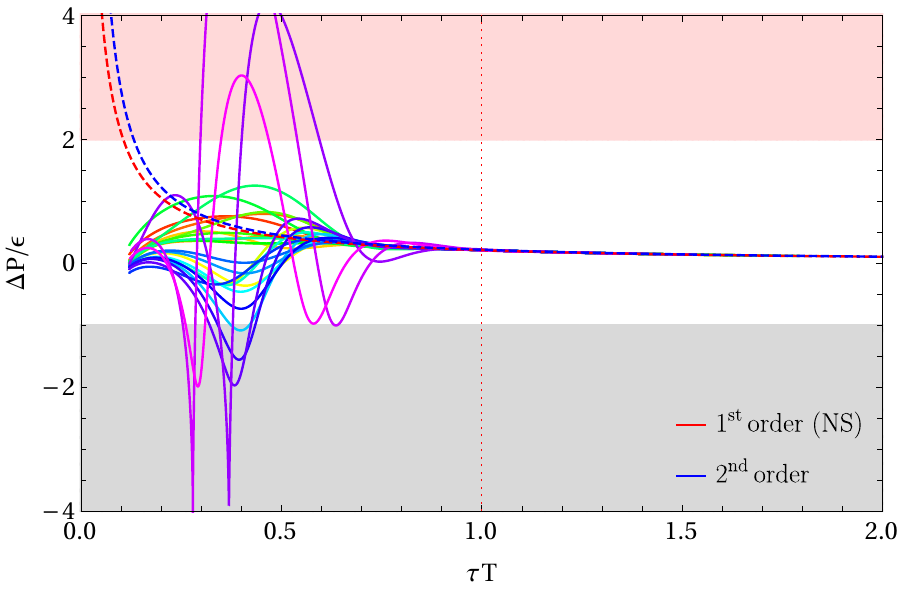}\hfill
    \includegraphics[width=0.49\textwidth]{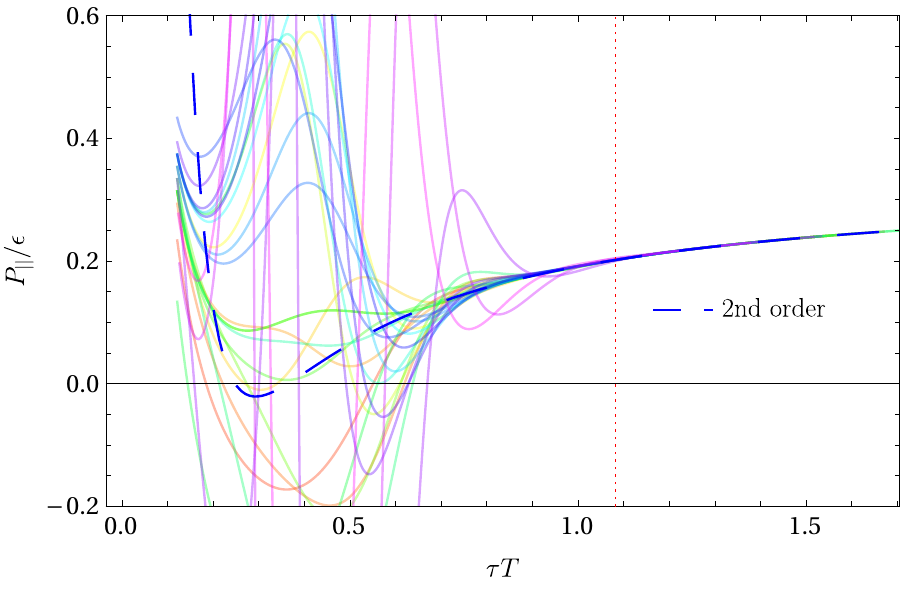}
    \caption{
    $\Delta P / \epsilon$ (left) and as a function of $
    \tau T$.
    The dashed curves are the first and second order solutions from Bjorken flow~\cite{Rougemont:2021qyk}.
    The red and gray areas indicate the regions where WEC and DEC are violated (red) and where only the DEC is violated 
    (gray)~\cite{Rougemont:2021qyk}.
    In some works not $\Delta P / \epsilon$ is investigated, but $P_{\para} / \epsilon$, so we also give a plot
    here (right). 
    }
    \label{figs:deltaPandPL}
\end{figure}
The work of~\cite{Rougemont:2021qyk} demonstrated that reasonable $AdS$ geometries produced field theory energy-momentum tensors which violated the weak, dominant or both energy conditions. These two energy conditions are summarized as follows~\cite{Wald:1984rg,Carroll:2004st},
\begin{itemize}
    \item Weak Energy Condition (WEC): For any timelike vector $\psi^\alpha$ ($\psi^\alpha\psi^{\beta}g_{\alpha\beta}< 0$) the WEC states that the energy-momentum tensor must obey $T_{\alpha\beta}\psi^\alpha\psi^\beta \geq 0$ .
    \item Dominant Energy Condition (DEC): For an timelike vector $\psi^{\alpha}$ the DEC requires that spacetime vector $\chi$ defined as $\chi^\alpha=-\tensor{T}{^\alpha_\beta}\psi^\beta$ must be a future directed null or timelike vector ($\chi^{\alpha}\chi_\alpha \leq 0$). This is equivalent to the condition that $T_{\alpha\gamma}\tensor{T}{^\alpha_\lambda}\psi^\gamma\psi^\lambda \leq 0$. It is a trivial exercise to show that matter which obeys the DEC implies the matter also obeys the WEC. 
\end{itemize}
For a vacuum solution, $T_{\mu\nu}=0$ and both of the energy conditions are trivially satisfied. Hence we pause and note that an AdS geometry supported by reasonable matter (vacuum) may produce a state of the field theory, for which, the energy conditions as applied to the field theory are violated. As such, the energy conditions as applied to the state of the field theory may be used to place bounds on the initial gravitational data.

\end{document}